\documentclass[a4paper,english]{article}
\usepackage[T1]{fontenc}
\usepackage[latin1]{inputenc}
\usepackage{color}
\usepackage{amsmath}
\usepackage{amssymb}
\usepackage{graphicx}

\makeatletter

\providecommand{\tabularnewline}{\\}

\newcommand{\lyxaddress}[1]{
\par {\raggedright #1
\vspace{1.4em}
\noindent\par}
}

\@ifundefined{definecolor}
 {\usepackage{color}}{}
\@ifundefined{definecolor}
 {\@ifundefined{definecolor}
 {\usepackage{color}}{}
}{}
\usepackage{babel}

\usepackage{babel}\usepackage{amsfonts}\@ifundefined{definecolor}
 {\@ifundefined{definecolor}
 {\usepackage{color}}{}
}{}
\usepackage{array}\usepackage{supertabular}\usepackage{hhline}

\newcommand{\arraybslash}{\let\\\@arraycr}

\newcommand{\liststyleWWNumi}{%
\renewcommand\theenumi{\roman{enumi}}
\renewcommand\theenumii{\alph{enumii}}
\renewcommand\theenumiii{\roman{enumiii}}
\renewcommand\theenumiv{\arabic{enumiv}}
\renewcommand\labelenumi{(\theenumi)}
\renewcommand\labelenumii{\theenumii.}
\renewcommand\labelenumiii{\theenumiii.}
\renewcommand\labelenumiv{\theenumiv.}
}
\newcommand{\ps@Standard}{
  \renewcommand\@oddhead{}
  \renewcommand\@evenhead{}
  \renewcommand\@oddfoot{}
  \renewcommand\@evenfoot{}
  \renewcommand\thepage{\arabic{page}}
}

\title{}
\author{Paul Warburton}
\date{2015-03-13}

\usepackage{babel}

\newcommand{\average}[1]{\langle #1 \rangle}

\@ifundefined{showcaptionsetup}{}{%
 \PassOptionsToPackage{caption=false}{subfig}}


\usepackage{babel}
\usepackage[normalem]{ulem}

\makeatother

\usepackage{babel}
\begin{document}
\clearpage{}\setcounter{page}{1}\pagestyle{Standard}

\title{\textcolor{black}{Maximum-Entropy Inference with a Programmable Annealer}}

\author{\textcolor{black}{Nicholas Chancellor{*}'', Szilard Szoke\textasciicircum{},
Walter Vinci'$^{\dagger}$, Gabriel Aeppli$^{\ddagger}$ and Paul
A. Warburton''\textasciicircum{}}}

\maketitle

\lyxaddress{\textcolor{black}{``London Centre For Nanotechnology 19 Gordon St,
London UK WC1H 0AH}\\
\textcolor{black}{{} \textasciicircum{}Department of Electronic and
Electrical Engineering, UCL, Torrington Place London UK WC1E 7JE}\\
\textcolor{black}{{} ' University of Southern California Department
of Electrical Engineering 825 Bloom Walk Los Angeles CA. USA 90089
}\\
\textcolor{black}{{} $^{\dagger}$University of Southern California
Center for Quantum Information Science Technology 825 Bloom Walk Los
Angeles CA. USA 90089}\\
\textcolor{black}{{} $^{\ddagger}$Department of Physics, ETH Zürich,
CH-8093 Zürich, Switzerland}\\
\textcolor{black}{{} Department of Physics, École Polytechnique Fédérale
de Lausanne (EPFL), CH-1015 Lausanne, Switzerland}\\
\textcolor{black}{{} Synchrotron and Nanotechnology Department, Paul
Scherrer Institute, CH-5232, Villigen, Switzerland}\\
\textcolor{black}{{*}corresponding author email: nicholas.chancellor@gmail.com}\\
\textcolor{black}{{} }}
\begin{abstract}
\textcolor{black}{Optimisation problems typically involve finding
the ground state (}\textit{\textcolor{black}{i.e.}}\textcolor{black}{{}
the minimum energy configuration) of a cost function with respect
to many variables. If the variables are corrupted by noise then this
maximises the likelihood that the solution is correct. The maximum
entropy solution on the other hand takes the form of a Boltzmann distribution
over the ground and excited states of the cost function to correct
for noise. Here we use a programmable annealer for the information
decoding problem which we simulate as a random Ising model in a field.
We show experimentally that finite temperature maximum entropy decoding
can give slightly better bit-error-rates than the maximum likelihood
approach, confirming that useful information can be extracted from
the excited states of the annealer. Furthermore we introduce a bit-by-bit
analytical method which is agnostic to the specific application and
use it to show that the annealer samples from a highly Boltzmann-like
distribution. Machines of this kind are therefore candidates for use
in a variety of machine learning applications which exploit maximum
entropy inference, including language processing and image recognition.
We discuss possible applications to probing equilibration and probing
the performance of the quantum annealing algorithm. }
\end{abstract}

\section*{\textcolor{black}{Introduction}}

\subsection*{\textcolor{black}{Maximum Entropy Decoding}}

\textcolor{black}{A universal problem in science and engineering and
especially machine learning is to draw an objective conclusion from
measurements which are incomplete and/or corrupted by noise. It has
been long recognised \cite{key-1,key-2,key-3} that there are two
generic approaches for doing this:}

\textcolor{black}{\liststyleWWNumi }
\begin{enumerate}
\item \textcolor{black}{maximum a priori (MAP) estimation, which results
in a unique conclusion which maximises the likelihood of being correct; }
\item \textcolor{black}{marginal posterior maximisation (MPM), which results
in a probabilistic conclusion whose distribution maximises entropy. }
\end{enumerate}
\textcolor{black}{The maximum entropy MPM approach is an implementation
of Bayesian inference since prior knowledge about the expected distribution
of the measurements is required. Maximum entropy modeling has been
shown to be a highly effective tool for solving problems in diverse
fields including computational linguistics \cite{key-4}, unsupervised
machine learning \cite{key-5}, independent component analysis \cite{key-6},
ecological modeling \cite{key-7}, genetics \cite{key-8}, astrophysics
\cite{key-9}, solid state materials physics \cite{key-10} and financial
analytics\cite{Mistrulli(2011)}.}

\textcolor{black}{As an archetypal example of Bayesian inference we
consider the decoding problem, and specifically the extent to which
the maximum entropy MPM method outperforms the maximum likelihood
MAP method when implemented on a programmable Josephson junction annealer.
Decoding is a canonical, computationally hard optimisation problem.
The goal is to extract information from a signal which has been corrupted
by noise on a transmission channel. By adding }\textit{\textcolor{black}{M}}\textcolor{black}{{}
redundant bits to the }\textit{\textcolor{black}{N}}\textcolor{black}{{}
transmitted information bits it is possible to recover the data exactly
provided that }\textit{\textcolor{black}{M}}\textcolor{black}{/}\textit{\textcolor{black}{N}}\textcolor{black}{{}
exceed a noise-dependent threshold value set by Shannon's theorem.
Turbo codes and low-density parity check codes represent the current
state-of-the-art for decoding and come close to reaching the Shannon
limit. We choose the decoding problem since, as we show below, it
maps directly onto a hardware implementation of the Ising spin glass
and the prior knowledge required to maximise the entropy of the decoding
information is precisely quantifiable and described by a single parameter.}

\subsection*{\textcolor{black}{Ising Code}}

\textcolor{black}{Sourlas \cite{key-11} noted the algebraic similarity
between parity check bits in information coding and magnetically-coupled
spins in an Ising spin glass. He introduced the Ising code, in which
}\textit{\textcolor{black}{N}}\textcolor{black}{{} information bits
$X_{i}\in\{0,1\}$ are mapped onto }\textit{\textcolor{black}{N}}\textcolor{black}{{}
spins $\sigma_{i}\in\{-1,1\}$. A set of $M>N$ couplers }\textbf{\textit{\textcolor{black}{J}}}\textcolor{black}{{}
can be defined from these bits such that $H=\sum_{i_{1},i_{2}\ldots i_{p}}C_{i_{1},i_{2}\ldots i_{p}}J_{i_{1},i_{2}\ldots i_{p}}\prod_{k=1}^{m}\sigma_{i_{k}}^{z}$
where $C_{\mbox{\ensuremath{i_{1}},\ensuremath{i_{2}\ldots i_{m}}}}\in\{0,1\}$
is the }\textcolor{black}{\emph{connectivity}}\textcolor{black}{{} of
an underlying graph and $m$ is the locality of the graph. The set
}\textbf{\textit{\textcolor{black}{J}}}\textcolor{black}{{} of couplers
is transmitted over a noisy channel. The maximum }\textit{\textcolor{black}{likelihood}}\textcolor{black}{{}
decoded word then corresponds to the ground state (}\textit{\textcolor{black}{i.e.
}}\textcolor{black}{zero temperature) spin configuration for the received
(corrupted) set of couplers $\mathbf{J'}$. For $M\gg N$ and large
$m$ the Ising code asymptotically approaches the Shannon limit.}

\textcolor{black}{The maximum }\textit{\textcolor{black}{entropy}}\textcolor{black}{{}
decoded word corresponds to the sign of the thermal average of the
state of each spin at a finite temperature $T_{ME}>0$:}

\textcolor{black}{
\begin{equation}
\sigma_{i}^{decoded}(T_{ME})\equiv\textrm{sgn}(\average{\sigma_{i}^{z}})=\textrm{sgn}\left(\sum_{k}\sigma_{k}^{z}\,\textrm{e}^{\frac{-E_{k}}{k_{B}\, T_{ME}}}\right),\label{eq:decode}
\end{equation}
}

\textcolor{black}{where $E_{k}$ is the energy of a state and $\sigma_{k}^{z}$
is its orientation (The usual normalisation term is omitted here since
it is always positive.) Rujan \cite{key-13} showed that}\textit{\textcolor{black}{{}
$T_{ME}$}}\textcolor{black}{{} is given by the so-called Nishimori
temperature \cite{key-14} which is a monotonic function of the magnitude
of the channel noise and is therefore a noise dependent quantity.
In other words, given prior information about the channel noise (and
therefore about the likely distribution of the received set of couplers
}\textbf{\textit{\textcolor{black}{J'}}}\textcolor{black}{), maximum
entropy inference can be used to obtain better decoding performance
than the maximum likelihood approach. More recently the role of quantum
fluctuations on inferential problems including decoding have been
considered \cite{Otsubo(2012),Otsubo(2014),Inoue(2015)}. The current
paper does not seek to determine the role of quantum fluctuations,
in finding an optimal solution; instead it considers the extent to
which thermal fluctuations (which will necessarily be present in any
real implementation of a Sourlas decoder) may be exploited.}

\subsection*{\textcolor{black}{Programmable Annealer}}

\textcolor{black}{The D-Wave chip \cite{key-15,key-16} is a superconducting
integrated circuit implementation of an Ising spin glass in a local
(longitudinal) random field, $H_{Ising}$, to which a local transverse
field can be applied to relax the spins \cite{Brooke1999}. In the
case where minimum energy (}\textit{\textcolor{black}{i.e. }}\textcolor{black}{maximum
likelihood) solutions are retained, its use in machine learning \cite{key-17,key-18}
and a number of other applications \cite{key-19,key-20,key-21,Boixo2014,Venturelli2014,Rieffel2015}
has been proposed and/or demonstrated. There is significantly less
work on exploiting excited states in computation. Their use was first
analyzed in the context of characterizing graph structure \cite{key-22}.
Subsequently,{} the error correction protocol in \cite{Pudenz2014,Pudenz2015},
used excited states to find the lowest energy state of an error-corrected
Hamiltonian; this cannot be considered a maximum entropy application.
It is also worth noting that \cite{Boixo2014} used distance of found
excited states from the ground state as a metric of success for the
D-Wave annealer. }

\textcolor{black}{A time-dependent Hamiltonian describes the chip:}

\textcolor{black}{
\begin{equation}
H(t)=-A(t)\sum_{i}\sigma_{i}^{x}+\alpha B(t)H_{Ising}\label{eq:Ht}
\end{equation}
}

\textcolor{black}{where}

\textcolor{black}{
\begin{equation}
H_{Ising}=-\sum_{i}h_{i}\sigma_{i}^{z}-\sum_{i,j\in\chi}J_{ij}\sigma_{i}^{z}\sigma_{j}^{z}\label{eq:ISGham}
\end{equation}
}

\begin{figure}
\begin{centering}
\textcolor{black}{\includegraphics[scale=0.2]{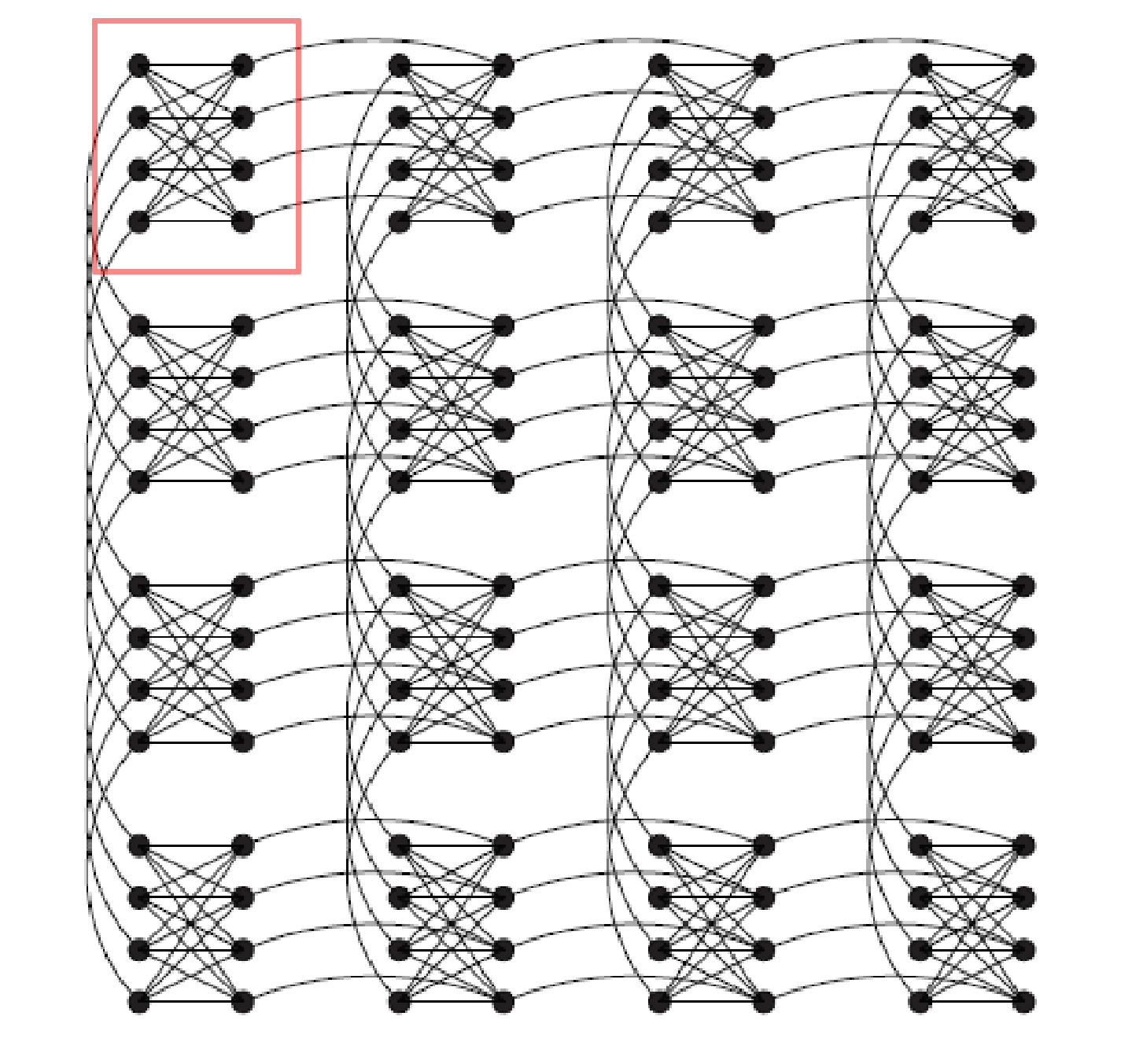} }
\par\end{centering}

\textcolor{black}{\caption{\label{fig:chimera4x4}Chimera graphs used for this study. The full
figure shows a 4x4 array of unit cells each containing eight spins,
a single example of which is shown in the rectangle at the top left
of the figure. Each dot represents a spin and each line a coupler.}
}
\end{figure}

\textcolor{black}{Here $\sigma^{x}$ and $\sigma^{z}$ are the Pauli
spin matrices and the }\textit{\textcolor{black}{$h_{i}$}}\textcolor{black}{{}
and $J_{ij}$ are user-programmable local fields and couplers respectively.
$\alpha$ sets the overall Ising energy scale and is also user-programmable.
The connectivity of the D-Wave machine is described by the so-called
Chimera graph, }\textit{\textcolor{black}{$\chi$}}\textcolor{black}{,
as shown in Fig. \ref{fig:chimera4x4}. The connectivity is sparse
with all couplers being two-local. Each spin is coupled to at most
either five or six other spins. The magnitude }\textit{\textcolor{black}{A}}\textcolor{black}{(}\textit{\textcolor{black}{t}}\textcolor{black}{)
of the transverse field sets the scale of quantum fluctuations on
the chip. During the course of a single optimisation run, }\textit{\textcolor{black}{A}}\textcolor{black}{(}\textit{\textcolor{black}{t}}\textcolor{black}{)
is adiabatically reduced to near zero in a manner analogous to simulated
annealing in which the scale of thermal fluctuations (}\textit{\textcolor{black}{i.e.}}\textcolor{black}{{}
the temperature) is reduced to zero. At $t=t_{f}$ the system is described
by $H(t)\approx\alpha B(t)\, H_{Ising}$ and the dynamics are fully
classical. If the chip were operated at zero temperature then, in
the absence of non-adiabatic transitions and control errors, at $t=t_{f}$
, the spins would be in the ground state configuration of $H_{Ising}$.
However, since the chip is operated at finite temperature }\textit{\textcolor{black}{$T_{chip}$}}\textcolor{black}{,
there is a non-zero probability that excited states will be occupied
at $t=t_{f}$. The non-zero occupation of the excited states suggests
that the spin configuration which maximises the entropy of the system
differs from the mean orientation in the ground state configurations.}

\textcolor{black}{The facts that the D-Wave chip operates at finite
temperature and that the dynamics at the end of the annealing process
are fully classical have led to a debate about the extent to which
quantum mechanics plays any role in its computational output \cite{Boixo2014,Venturelli2014,Albash2015,Shin(2014),Kadowaki1998,Farhi2001,Santoro2002,Ronnow2014,Heim2015,Mandra2014,Kechedzhi2015,Hauke2015,Boixo(2015)}.
For the entropy to be maximised, however, it is only necessary that
the final distribution of excited states follows the Boltzmann distribution,
which is }\textcolor{black}{\emph{by construction }}\textcolor{black}{the
one which maximizes the entropy for a given total ensemble energy.
In this context the long timescale of the annealing process can be
seen as an advantage in that it allows the system to fully thermalise.
Nevertheless the transverse field term in equation \ref{eq:Ht} may
help prevent local minima from trapping the system as suggested in
\cite{Boixo(2015),Battaglia2006}.}

\textcolor{black}{}

\textcolor{black}{In our experiments we perform }\textit{\textcolor{black}{encoding}}\textcolor{black}{{}
with conventional computational resources. The gauge symmetry of the
Ising spin glass allows us without loss of generality only to consider
a data word $\textbf{\textit{X}}$ consisting of a string of $N$
'ones'. In the Ising code as described by Sourlas \cite{key-11} the
transmitted codeword consists of the }\textit{\textcolor{black}{M}}\textcolor{black}{{}
couplers (here all ferromagnetic, }\textit{\textcolor{black}{$J_{ij}=1$}}\textcolor{black}{).
Based on Eq. \ref{eq:decode} and \ref{eq:ISGham} we note however
that if $\textit{h}_{i}$ = 0 $\forall\textit{i}$ then for the noise
free transmission channel, there are two trivial uniform solutions
for the decoded value $\sigma_{i}^{decoded}(T)=+1$ or $\sigma_{i}^{decoded}(T)=-1$,
due to the overall $\mathbb{Z}_{2}$ symmetry. To break this symmetry,
in addition to transmitting the couplers over the noisy transmission
line, we also transmit the values of the local fields $h_{i}$, with
assigned values $h_{i}=2X_{i}-1$ what we are effectively doing is
sending both the N original information bits and the parity bits $J_{ij}$
over the noisy channel. At the coding stage the coupling graph is
selected to match the connectivity of the D-Wave chip for ease of
subsequent decoding. The rate and distance of the Ising code as implemented
on the Chimera graph depend upon the word length }\textit{\textcolor{black}{N}}\textcolor{black}{{}
(}\textit{\textcolor{black}{i.e.}}\textcolor{black}{{} upon the number
of spins) and are shown in tab. \ref{tab:mappings}.}

\textcolor{black}{}
\begin{table*}
\begin{centering}
\textcolor{black}{}%
\begin{tabular}{|c|c|c|c|c|}
\hline 
\textcolor{black}{Word length, }\textit{\textcolor{black}{N}}\textcolor{black}{{} } & \textcolor{black}{Chimera mapping } & \textcolor{black}{no. of couplers, $M$ } & \textcolor{black}{rate of code }\textit{\textcolor{black}{N}}\textcolor{black}{/(}\textit{\textcolor{black}{M}}\textcolor{black}{{}
+ }\textit{\textcolor{black}{N}}\textcolor{black}{) } & \textcolor{black}{Distance of code}\tabularnewline
\hline 
\hline 
\textcolor{black}{8 } & \textcolor{black}{Single unit cell } & \textcolor{black}{16 } & \textcolor{black}{$0.\bar{3}$ } & \textcolor{black}{5}\tabularnewline
\hline 
\textcolor{black}{128 } & \textcolor{black}{4x4 array of unit cells } & \textcolor{black}{352 } & \textcolor{black}{0.276 } & \textcolor{black}{6}\tabularnewline
\hline 
\end{tabular}
\par\end{centering}

\textcolor{black}{\caption{\label{tab:mappings} Table of relevant quantities for different mappings.}
}
\end{table*}

\textcolor{black}{We model transmission over the binary symmetric
channel (BSC). Here the probability that any local field is received
corrupted (}\textit{\textcolor{black}{i.e.}}\textcolor{black}{{} $h_{i}=-1$)
is equal to the probability that any coupler is received corrupted
(}\textit{\textcolor{black}{i.e.}}\textcolor{black}{{} anti-ferromagnetic,
$J_{ij}=-1$) and given by the so-called crossover probability }\textit{\textcolor{black}{p}}\textcolor{black}{.
Corrupted bits and couplers are uncorrelated. For this channel the
Nishimori temperature is}

\textcolor{black}{
\begin{equation}
T_{Nish}=2\,(\ln(\frac{1-p}{p}))^{-1}.
\end{equation}
}

\textit{\textcolor{black}{Decoding}}\textcolor{black}{{} is performed
by the D-Wave chip according to the annealing schedule which begins
with $A(t=0)\gg B(t=0)$ and ends with $B(t=t_{f})\gg A(t=t_{f})$
with $t_{f}=20\mu s$. Here }\textit{\textcolor{black}{$H_{Ising}$}}\textcolor{black}{{}
is defined by the bias fields $h^{'}\in\{-1,1\}$ (corresponding to
the received corrupted set of }\textit{\textcolor{black}{N}}\textcolor{black}{{}
data bits) and the couplers $J'_{ij}\in\{-1,1\}$ (corresponding to
the received corrupted set of }\textit{\textcolor{black}{M}}\textcolor{black}{{}
couplers). Finite temperature decoding consists of repeating the annealing
process many times to find the sign of the average of each spin:}

\textcolor{black}{
\begin{equation}
\sigma_{i}^{decoded}\equiv\textrm{sgn}(\average{\sigma_{i}^{z}})=\textrm{sgn}\left(\sum_{k}\sigma_{i,k}^{z}n_{k}\right),\label{eq:decoding}
\end{equation}
}

\textcolor{black}{where }\textit{\textcolor{black}{$n_{k}$}}\textcolor{black}{{}
is the number of occasions that state }\textit{\textcolor{black}{k}}\textcolor{black}{{}
is occupied at the end of the anneal and $\sigma_{i,k}^{z}\in\{\pm1\}$
the orientation of spin $i$ in state $k$. Under the assumption that
the D-Wave machine samples from a Boltzmann distribution, one could
in principle adjust its temperature ${T_{chip}}$ to maximise entropy.
However since the machine is only calibrated at a fixed temperature
(set by the base temperature of the dilution refrigerator), we instead
vary the Nishimori temperature at the encoding stage and keep the
machine temperature constant. Maximum likelihood decoding is the special
case of equation \ref{eq:decoding} where only the spin configurations
corresponding to the ground state (or multiple degenerate ground states)
are considered.}

\textcolor{black}{For small systems we can compare the decoding performance
of the D-Wave machine with an analytical model. In this case the spectrum
of states corresponding to the received set of bias fields and couplers
is calculated by summing exhaustively over all states; the population
of these states is assumed to follow a Boltzmann distribution. We
further define{} temperatures at which the value of $\sigma_{i}^{decoded}$
in Eq. \ref{eq:decode} changes, which we call }\textcolor{black}{\emph{spin-sign
transition}}\textcolor{black}{{} temperatures. These transitions allow
us to not only estimate the temperature which best fits the experimental
data, but also to determine to what extent our data matches a Boltzmann
distribution, which can be treated as a method of measuring equilibration.}

\subsection*{\textcolor{black}{Connection to spin glasses}}

\textcolor{black}{It is worth asking questions about the Hamiltonians
we use based on spin glass theory, which can be used as a way of gauging
the 'hardness' of solving a typical instance. A spin glass phase which
exists at finite temperature implies that finding the ground state
is 'hard' using realistic (i.e. local) thermal fluctuations or software
algorithms such as Monte Carlo and parallel tempering. It is important
to note that this statement is in one sense stronger than the statement
that a problem is NP-hard, because it is a statement about typical
rather than worst-case hardness, but weaker in another sense because
it only addresses one approach to solving the problem. It is worth
noting additionally the recent propolsal for a general purpose algorithm
which has been shown to be very efficeint for solving typical instances
from classes of Hamiltonians with zero temperature spin glass transitions,
such as the chimera graph without fields \cite{Zhu2015}.}

\textcolor{black}{There have not been any studies of whether a Chimera
graph in a field has a finite temperature spin glass phase. Zero field
random Chimera graphs have been shown to not have a finite temperature
spin glass transition\cite{Katzgraber2014}. There are many other
examples however where a model has a finite temperature spin glass
phase without a field, but the inclusion of a field causes this phase
to vanish \cite{Bray1980,Stauffer1978}. The 3D Ising Edward-Anderson
model provides a concrete example of this phenomenon \cite{Young2004,Feng2014}.
This, coupled with the numerical evidence that chimera graphs which
are scaled up with fixed unit cell length will have the thermodynamic
properties of a 2 dimensional graph in the large system limit \cite{Katzgraber2014,Weigel2014}
suggest that a chimera graph in field will not have a finite temperature
spin glass transition. Although a specific study is absent for the
Chimera graph in a field, there is significant evidence to suggest
that a finite temperature spin glass transition is unlikely to be
present. }

\textcolor{black}{It is important to note that the{} set of Hamiltonians
which we study here is }\textcolor{black}{\emph{not }}\textcolor{black}{what
is traditionally defined as a Random Field Ising Model (RFIM). RFIMs
have the additional constraint that all $J_{ij}\geq0$. This subtle
distinction is important, because it has been shown that RFIMs do
not have a spin glass phase in equilibrium \cite{Krzakala2010,Krzakala2011}.
As can be done with any 2-body interaction Hamiltonian, our Hamiltonians
can all be defined in a gauge where all of the local fields are in
the same direction, and act as an effective uniform global field.
They therefore can be considered a hybrid of a Mattis Spin glass \cite{Mattis(1976)}
and a ``spin glass in a uniform field'', approaching the latter
exactly when $p\rightarrow\frac{1}{2}$. While there are open questions
about the nature of the spin glass transition for a spin glass in
a uniform field, it is generally accepted that (for some topologies)
these models do still behave as spin glasses. In particular, it is
an open question\cite{Larson2013} whether Replica Symmetry Breaking\cite{Parisi1980}
(RSB) provides an appropriate description of these models, or whether
they are better described in a phenomenological ``droplet'' picture\cite{Fisher1987,Fisher1988}.
As for quantum annealing itself, the most relevant experiments have
been performed in LiHoxY1-xF4, where the transverse field imposed
in the laboratory introduces not only quantum fluctuations, but also
internal random longitudinal fields\cite{Silevich(2007),Tabei(2006),Schechter(2008)}.}

\textcolor{black}{The techniques laid out in this manuscript provide
a way of probing equilibration. If{} the spin-sign transition temperatures{}
(where the value of Eq. \ref{eq:decode} changes) which we calculate
using exhaustive summing and tree decomposition methods could be calculated
using a method which supports larger system sizes, and an interesting
experimental or numerical system of such a size were available{}
then our methods could prove a valuable tool to examine equilibration
. It is also worth pointing out that, while we choose to examine spin-sign
transitions due to the connection with the real world problem of decoding,
an analogue of Eq. \ref{eq:decode} for correlations, ${\color{red}{\color{black}\textrm{sgn}\left(\sum_{k}\sigma_{i,k}^{z}\sigma_{j,k}^{z}\textrm{e}^{\frac{-E_{k}}{k_{B}\, T_{ME}}}\right),}}$
and correlation-sign transitions could be defined in analogue to our
spin-sign transitions. While spin-sign is only meaningful in models
where the $\mathbb{Z}_{2}$ symmetry of the Ising model is broken,
correlation-sign transitions carry no such restriction, and could
be used with Hamiltonians with no field terms.}

\textcolor{black}{Even without having access to a method which can
reliably calculate spin- or correlation- sign transitions for larger
systems, generalizations of our methods could be interesting for example
to examine whether equilibrium or non-equilibrium effects are the
limiting factor in small instances of various benchmarking experiments,
such as those performed in \cite{Hen2015}.}

\subsection*{\textcolor{black}{Experimental methods}}

\textcolor{black}{All decoding experiments were performed on the D-Wave
Vesuvius processor located at the Information Sciences Institute of
the University of Southern California. This processor contains 512
bits in an 8x8 array of unit cells. Due to fabrication errors, nine
of the bits fall outside of the acceptable calibration range. Any
unit cells containing one or more of these uncalibrated bits were
therefore not used in our experiments. The annealing time $t_{f}$
was fixed at $20\mu s$. Each bit in each received corrupted word
was mapped onto a randomly chosen gauge on the chip. To determine
the bit-error-rate for a given value of crossover probability,{}
we first sum the observed{} values of the orientation of each of
the spins used and take the sign using the fact that Eq. \ref{eq:decode}{}
becomes{} $\textrm{sgn}\left(\sum_{k}\sigma_{k}^{z}\,\delta(E_{k}-\min(E_{k})\right)$
in the limit of $T\rightarrow0${} . We then define the bit-error-rate
as the probability with which the decoded spin orientation disagrees
with the original message. Because we have chosen a trivial message,
this amounts to calculating $r_{n}=\frac{1}{N}\sum_{i}\frac{1}{2}(1-\textrm{sgn}(\average{\sigma_{i}^{z}}))$
for each Hamiltonian $H{}_{n}$ and performing a sum weighted by the
probability $q(H_{n},p)$ that this Hamiltonian will be generated.
For a given crossover probability $p$, the bit-error-rate is given
by}

\textcolor{black}{
\begin{equation}
r_{tot}(p)=\sum_{n=1}^{2^{N+M}}q(H_{n},p)\, r_{n}.
\end{equation}
}

\textcolor{black}{In practice, we simplify this calculation greatly
by taking advantage of the fact that the value of $q(H_{n},p)=\frac{1}{2^{N}}(p)^{N_{corr}}(1-p)^{N+M-N_{corr}}\tbinom{N+M}{N_{corr}}$,
only depends on the number of couplers and fields which are corrupted,
$N_{corr}$, and the crossover probability $p$. We explain how we
exploit this fact in Sec. 1 of the supplemental material.}

\subsection*{\textcolor{black}{Macroscopic Analysis of Decoding}}

\subsubsection*{\textcolor{black}{Single unit cell}}

\begin{figure}
\begin{centering}
\textcolor{black}{\includegraphics[scale=0.5]{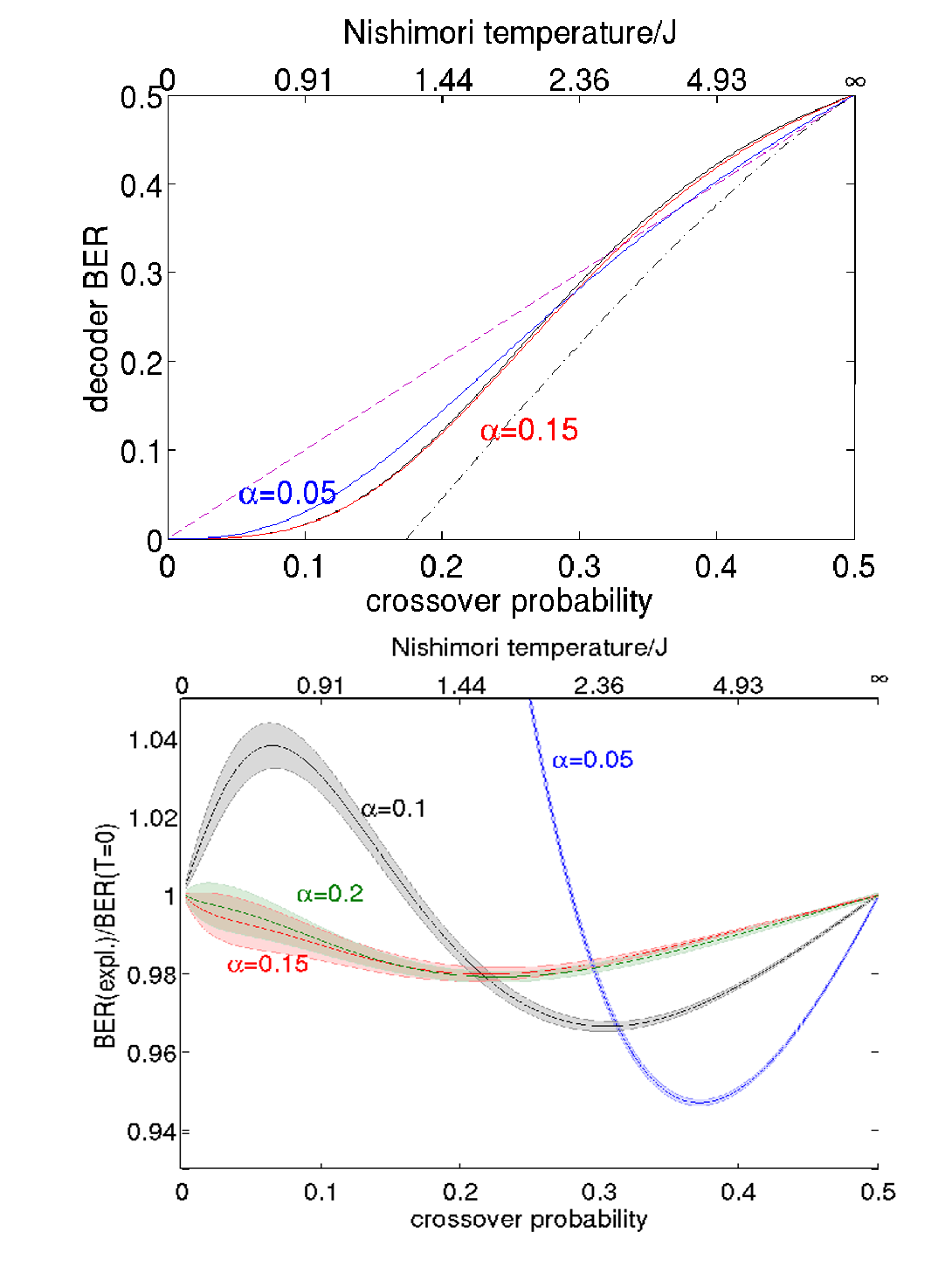} }
\par\end{centering}

\textcolor{black}{\caption{\label{fig:decoderBER_sing}Top: Bit-error-rate \textcolor{black}{(BER)}
for a single Chimera unit cells plotted as a function of channel crossover
probability. The solid line black line represents theoretically calculated
ground state decoding. Red line is experimental data with $\alpha=0.15$
which corresponds to a coupling energy scale of $13.4$ mK (at the
point when the dynamics freeze\textcolor{black}{, this energy scale
was established based on freezing when the transverse field energy
scale is $0.1$GHz as suggested in private communication with Mohammad
Amin.}) which is the order of the base temperature $17$ mK of the
cryostat. Blue line is experimental data with $\alpha=0.05$ which
corresponds to a coupling energy scale of $4.99$ mK. The dot-dash
line shows the Shannon-limited minimum achievable bit-error-rate for
a decoder of rate $0.333$, corresponding to the rate of the Ising
code on a single unit-cell. The dashed line is a guide to the eye
showing the locus of points where the decoder bit-error-rate is equal
to the crossover probability. Bottom: Ratio of experimental decoding
\textcolor{black}{error rate to maximum likelihood decoding error
rate.} Shaded areas are one standard deviation estimated by bootstrapping.
$\alpha=0.1$ corresponds to $8.9$ mK and $\alpha=0.2$ corresponds
to $17.8$ mK. }
}
\end{figure}

\textcolor{black}{The connectivity is defined by a single Chimera
unit cell containing 8 spins. Hence the number of data bits is }\textit{\textcolor{black}{N}}\textcolor{black}{{}
= 8 and the number of couplers is }\textit{\textcolor{black}{M}}\textcolor{black}{{}
= 16. The decoded bit error rate is plotted as a function of the crossover
probability, }\textit{\textcolor{black}{p}}\textcolor{black}{{} (or
equivalently the Nishimori temperature, $T_{Nish}$) in the top subfigure
of Fig. \ref{fig:decoderBER_sing}. The decoder is useful (}\textit{\textcolor{black}{i.e.}}\textcolor{black}{{}
the decoded bit error rate is lower than }\textit{\textcolor{black}{p}}\textcolor{black}{)
for }\textit{\textcolor{black}{p}}\textcolor{black}{{} {\textless}
0.327 but compares unfavourably with the Shannon-limited performance.}

\textcolor{black}{}

\textcolor{black}{We now consider whether the maximum entropy approach
is effective in reducing the bit error rate. The bottom subgraph of
Fig. \ref{fig:decoderBER_sing} shows the ratio of the experimental
to theoretical bit-error-rates for zero temperature.We mimic the effect
of changing the decoding temperature by changing the energy scale
}\textit{\textcolor{black}{${\alpha}$}}\textcolor{black}{{} of the
Ising Hamiltonian in Eq. \ref{eq:Ht}; here higher values of }\textit{\textcolor{black}{${\alpha}$}}\textcolor{black}{{}
correspond to lower decoding temperatures. A similar technique has
been applied in \cite{Albash2015} in the context of identifying quantum
effects. At }\textit{\textcolor{black}{${\alpha}$}}\textcolor{black}{{}
= 0.05 and }\textit{\textcolor{black}{${\alpha}$}}\textcolor{black}{{}
= 0.1 the D-Wave finite temperature decoding compares poorly with
ground state decoding at low noise levels but is superior at higher
values of the crossover probability. As the effective decoding temperature
is lowered (}\textit{\textcolor{black}{e.g. ${\alpha}$}}\textcolor{black}{{}
= 0.2) we reach the regime where finite temperature decoding always
outperforms maximum likelihood decoding. Further reducing the decoding
temperature has no effect on the bit-error-rate for reasons which
we will explain later.}

\textcolor{black}{It is worth noting that the maximum entropy method,
which relies simply on measuring thermal expectation values, never
outperforms the maximum likelihood ``ground state'' calculations
by more than 10\%. This most likely is a reflection of the extent
to which thermal fluctuations for a quadratic Ising model can mimic
the effects of higher order spin couplings (i.e. beyond quadratic)
which optimized Sourlas codes for error correction contain.}

\begin{figure}
\begin{centering}
\textcolor{black}{\includegraphics[scale=0.5]{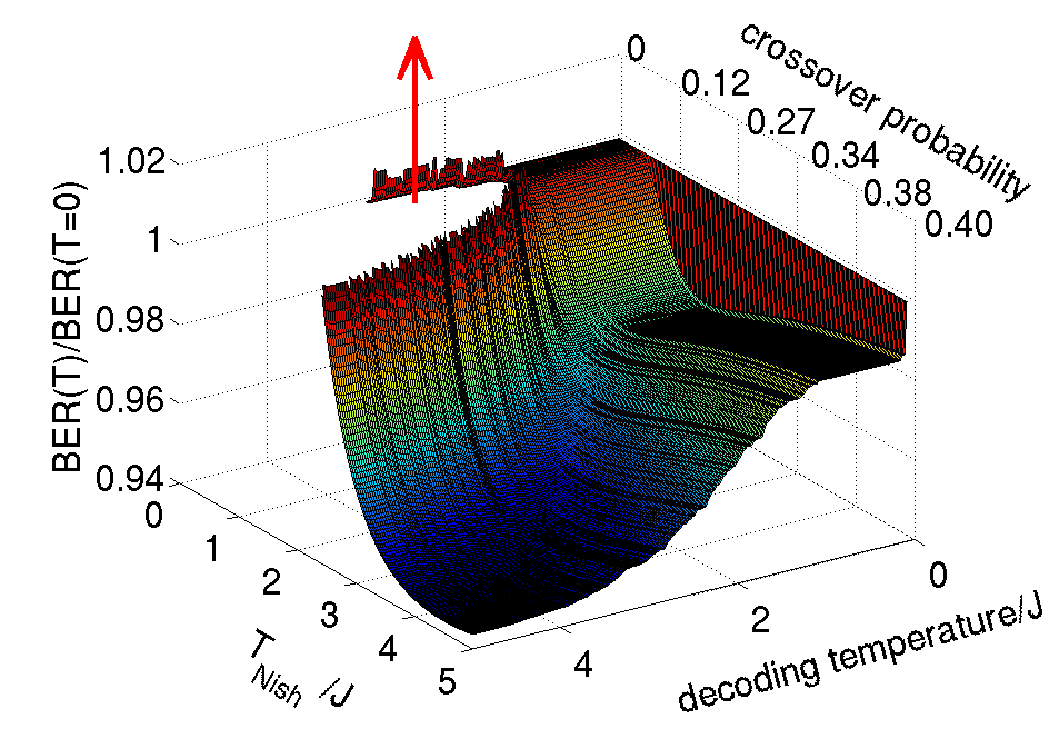} }
\par\end{centering}

\textcolor{black}{\caption{\label{fig:decoding3D}Analytically calculated bit-error-rate \textcolor{black}{(BER)
}for a single unit-cell of the Chimera graph, plotted as a function
of the decoding temperature and the Nishimori temperature. The bit-error-rates
are normalised with respect to those obtained using maximum likelihood
decoding. }
}
\end{figure}

\textcolor{black}{We now compare the finite temperature experimental
result of Fig. \ref{fig:decoderBER_sing} with the analytical result
obtained by exhaustive sums assuming a Boltzmann distribution. At
low Nishimori temperature (}\textit{\textcolor{black}{i.e.}}\textcolor{black}{{}
low channel noise) the Bayesian approach fares badly by comparison
with the maximum likelihood approach. In this regime the correct error-free
codeword is typically the closest allowed codeword (in the Hamming
sense) to the received corrupted codeword, and the error-free decoding
corresponds to the ground state spin configuration. As the channel
noise increases, however, the maximum entropy approach outperforms
ground state decoding for the quadratic Hamiltonian used here. It
can be seen therefore that, at least qualitatively, the D-Wave machine
behaves as a maximum entropy decoder. It has been proven \cite{key-14,Nishimori2001}that
the bit error rate is minimised when the decoding temperature is equal
to the Nishimori temperature $T=T_{Nish}$. Note however that owing
to the fact that this curve is discontinuous as a function of $T$,
the converse is not necessarily true - $i.e.$ for a fixed value of
$T$ it is }\textcolor{black}{\emph{not}}\textcolor{black}{{} necessarily
true that the optimum will be at $T_{Nish}=T$. See Sec. 2 of the
supplemental material accompanying this document for a graphical demonstration
that the results from \cite{key-14,Nishimori2001} do in fact hold
for the data in Fig. \ref{fig:decoding3D}.}

\textcolor{black}{}

\textcolor{black}{The discontinuity of the bit error rate as a function
of decoding temperature results from }\textit{\textcolor{black}{\emph{spin-sign
transitions}}}\textcolor{black}{{} as described earlier. Of particular
note is the plateau at normalised temperatures below 1 unit, suggesting
that there are no such spin-sign transitions in this temperature range.
It is for this reason that the experimental data do not differ significantly
from each other for $\alpha\gtrsim0.2$. An analysis based on these
transitions follows later in the manusript. Simulations (not shown,
but see Sec. 3 of the supplemental material) indicate that the magnitude
of the discontinuities and the temperature width of the plateau both
reduce as the system size increases. The experimental data in Fig.
\ref{fig:decoderBER_sing} reproduce the analytical data rather well,
confirming that the population of the excited states at the end of
the annealing process is at least reasonably well described by a Boltzmann
distribution.}

\begin{figure}
\begin{centering}
\textcolor{black}{\includegraphics[scale=0.5]{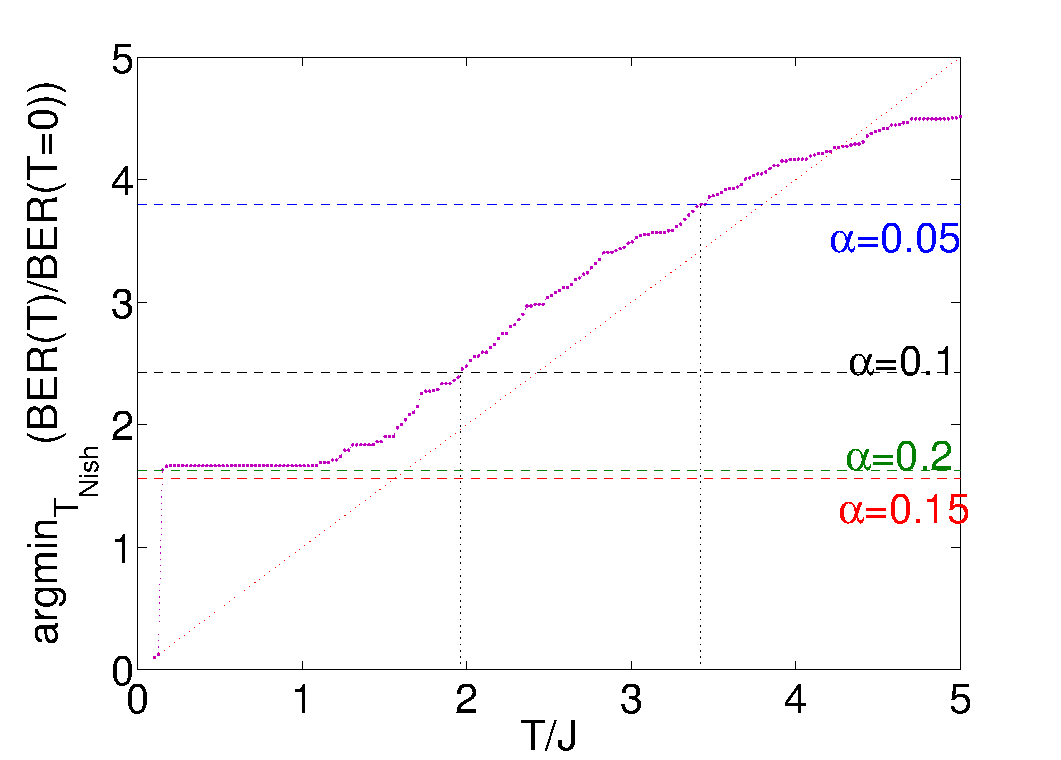} }
\par\end{centering}

\textcolor{black}{\caption{\label{fig:min_theory_experiment}The analytically calculated value
of the Nishimori temperature at which the bit-error-rate\textcolor{black}{{}
(BER) }ratio is minimised (i.e. entropy is maximised),plotted as a
function of the decoding temperature for a single Chimera unit cell.
The dotted line is included as a guide to the eye. Note that the data
points at $T/J<0.2$ are a numerical artifact caused by finite machine
precision. \textcolor{black}{The horizontal dashed lines show temperatures
obtained by numerically extracting the minima from the experimental
results from the annealing device in figure \ref{fig:decoderBER_sing}.
As explained in Sec. S1, these curves are given by a polynomial, the
minimum can therefore be found using a standard algorithm.} }
}
\end{figure}

\textcolor{black}{We further perform a direct comparison between the
theoretical data in Fig. \ref{fig:decoding3D} and the experimental
data in Fig. \ref{fig:decoderBER_sing}. Specifically, we consider
the value of $T_{Nish}$ at which the ratio of the maximum entropy
to maximum likelihood error rates is the smallest. Fig. \ref{fig:min_theory_experiment}
compares the experimentally observed minima to the theoretical predictions.
For $\alpha=0.2$ and $\alpha=0.15$ the predicted minima are within
a large 'plateau' and therefore the location of the minimum is well
explained by a wide range of temperatures. For $\alpha=0.1$ and $\alpha=0.05$
however we can make such an estimate, which puts the ``effective
temperature'' for $\alpha=0.1$ at around $T_{chip}/J=2$ and the
value for $\alpha=0.05$ at around $T_{chip}/J=3.5$. We later perform
analysis which allows us to make a much more accurate estimate of
these temperatures.}

\textcolor{black}{}

\subsection*{\textcolor{black}{4x4 Array of Unit Cells}}

\begin{figure}
\begin{centering}
\textcolor{black}{\includegraphics[scale=0.5]{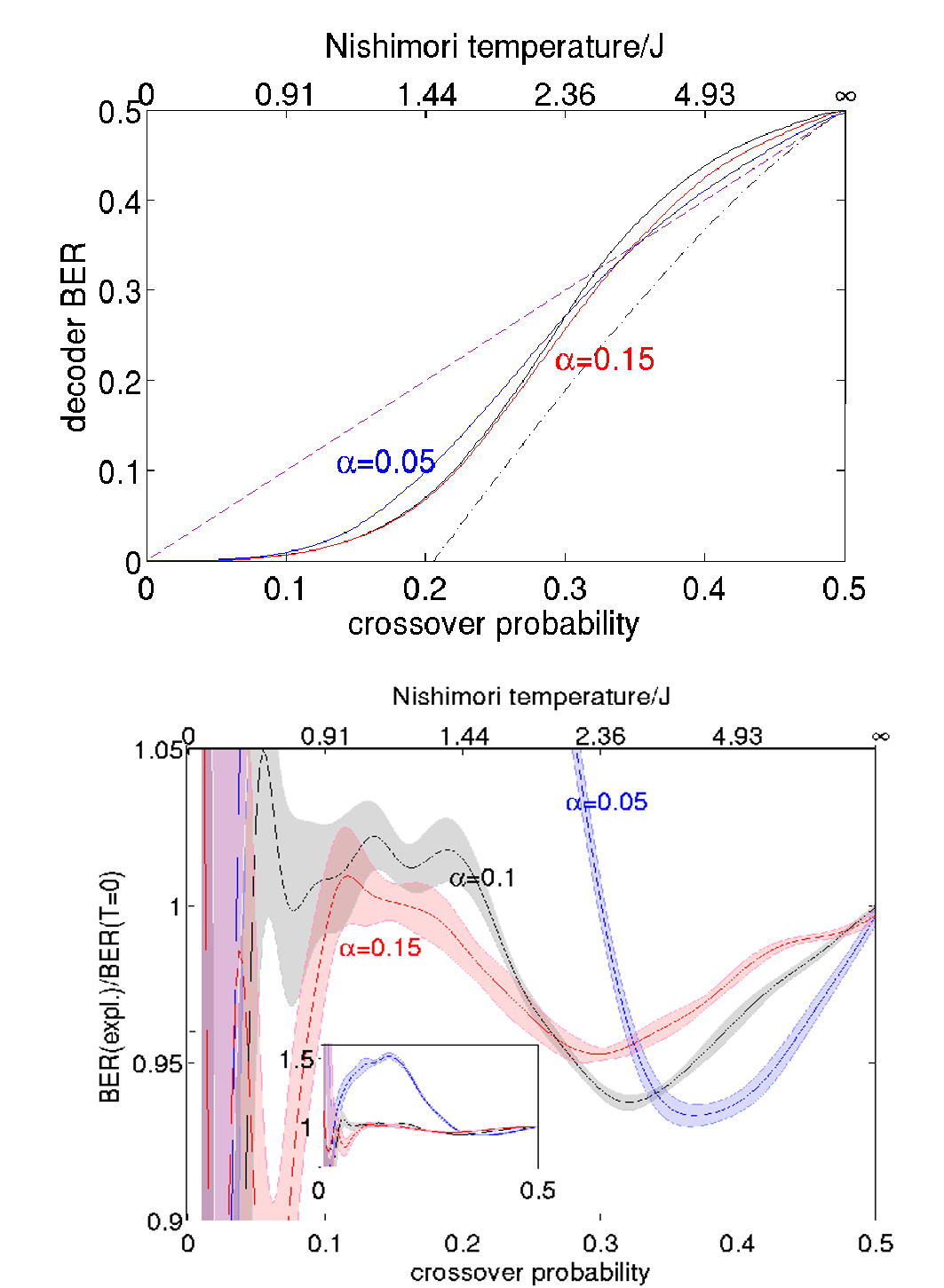} }
\par\end{centering}

\textcolor{black}{\caption{\label{fig:decoderBER_4x4}Top: Bit-error-rate \textcolor{black}{(BER)
}for a 4x4 array of Chimera unit cells plotted as a function of channel
crossover probability. The solid line black line represents theoretically
calculated ground state decoding for the 4x4 array. Red line represents
experimental data with $\alpha=0.15$ which corresponds to a coupling
energy scale of $13.4$ mK (at the point when the dynamics freeze)
which is the order of the base temperature $17$ mK of the cryostat.
Blue line represents experimental data with $\alpha=0.05$ which corresponds
to a coupling energy scale of $4.99$ mK. The dot-dash line shows
the Shannon-limited minimum achievable bit-error-rate for a decoder
of rate $0.276$, corresponding to the rate of the Ising code on a
4x4 chimera graph. The dashed line is a guide to the eye showing the
locus of points where the decoder bit-error-rate is equal to the crossover
probability. Bottom: Ratio of experimental decoding rate to maximum
likelihood decoding rate. Shaded areas are one standard deviation
estimated by bootstrapping. $\alpha=0.1$ corresponds to $8.9$ mK.}
}
\end{figure}

\textcolor{black}{To demonstrate that the maximum entropy approach
is effective for larger systems, let us now consider decoding for
a 4x4 array of Chimera unit cells. These data, with 128 information
bits and 352 couplers using the D-Wave chip are shown in Fig. \ref{fig:decoderBER_4x4}.
In this example the qualitative features of the analytical result
for an 8-bit word are reproduced. For each value of $\alpha$ there
is a wide range of crossover{} probabilities where the experimental
system outperforms exact zero temperature calculations. The feature
where a low $\alpha$ system performs very poorly compared to the
zero temperature result is also qualitatively reproduced.}

\textcolor{black}{It is worth noting that the difference between the
maximum entropy approach and the maximum likelihood approach is significantly
more dramatic for $\alpha=0.15$ in this case than it was for the
single unit cell. This may partially be due to the fact that the rate
of this code is lower, but is also probably related to the fact that
the larger system with more complex topology allows for a greater
richness of possible thermal fluctuations.}

\textcolor{black}{}

\subsection*{\textcolor{black}{Microscopic Analysis}}

\textcolor{black}{We have established that the macroscopic behavior
of spin decoding on our experimental system is quite similar to what
would be expected for a thermal distribution. While the macroscopic
investigation a demonstrates that the chip can be used for finite
temperature decoding, it does not provide strong evidence on how suitable
the chip may be for other maximum entropy tasks. To answer this question
we need to examine whether or not the individual spin orientations
on the annealing machine look similar to those expected from a Boltzmann
distribution.}

\textcolor{black}{The Boltzmann distribution is uniquely and by construction
that which maximizes entropy for a given energy cost. In the case
of maximum entropy tasks we can think of this as optimizing robustness
to uncertainty, given that we are willing to sacrifice a given amount
of optimality in our solution. The closer our result is to Boltzmann,
the closer the result is to optimizing this tradeoff. In the next
section we introduce the theory behind our microscopic analysis, and
apply it to our experimental data in the section following that.}

\textcolor{black}{It is important to note that, unlike the previous
section, the microscopic analysis is independent of the intended application.
We have chosen a decoding example for this study because the mapping
onto the D-Wave chip is trivial. The results of this section however
apply equally well for other, technologically useful maximum entropy
applications.}

\subsection*{\textcolor{black}{Microscopic theory of Spin-Sign transitions}}

\textcolor{black}{To examine the performance for maximum entropy tasks
on a microscopic scale, we must first have a theoretical understanding
of what is expected for a Boltzmann distribution. For a system which
consists of a finite number of spins the value of $\sigma_{i}^{\textrm{decoded}}(T)\in\{-1,1\}$
can only change a finite number of times, $n_{i}^{\textrm{trans}}$,
on the interval $T\in(0,\infty]$. For this reason we can uniquely
characterize the decoding of each spin by the set of temperatures
$\{T_{trans}\}_{i}$ at which the value of $\sigma_{i}^{\textrm{decoded}}(T)$
changes, and the value of $\sigma_{i}^{\textrm{decoded}}(T')$ at
any single temperature $T'\notin\{T_{trans}\}_{i}$. We will refer
to each of these temperature dependent changes in orientation as a
}\textcolor{black}{\emph{spin-sign transition.}}\textcolor{black}{{}
It may often be the case that for a given Hamiltonian a spin has no
spin-sign transitions and therefore $\{T_{trans}\}_{i}=\{\}$. For
these spins $\sigma_{i}^{\textrm{decoded}}(T)$ is independent of
$T$ and maximum entropy decoding can be considered trivial. There
can further be Hamiltonians in which this is the case for all spins,
for example this must be true for any Hamiltonian in which all couplers
and fields can be satisfied simultaneously.}

\textcolor{black}{To analyze the performance for a maximum entropy
application, we choose to focus on the non-trivial cases of spins
which have at least one spin-sign transition. Let us define the low
temperature decoding of a spin as $\sigma_{i}^{\textrm{low}}\equiv\lim_{\epsilon\rightarrow0}\sigma_{i}^{\textrm{decoded}}(\epsilon)$.
Note that we define this as a limit to avoid ambiguity in cases where
$\sigma_{i}^{\textrm{decoded}}(0)=0$. For an experimental system
we can define the probability of agreeing with the low temperature
result over many repetitions of the experiment $P_{i}^{\textrm{low}}\equiv P(\sigma_{n_{run}}^{\textrm{experiment}}=\sigma_{i}^{\textrm{low}})$
where $\sigma_{n_{run}}^{\textrm{experiment}}=\textrm{sgn}(\sum_{i=1}^{n_{run}}\sigma_{i}^{\textrm{experiment}})$
and $\sigma_{i}^{\textrm{experiment}}$ is the result of a single
experimental run. The probability of agreeing with the low temperature
result, $0\leq P_{i}^{\textrm{low}}\leq1$ takes a continuum of values
rather than just $0$ or $1$ because individual experiments do not
always yield the same results. We define $P_{i}^{\textrm{low}}$ in
this way for two reasons. First, the user interface to the device
places an upper limit on $n_{run}$, so we cannot be confident that
the result of a single experiment is statistically significant. Second
and more importantly, each experiment is subject to a set of control
errors, which do not change significantly during the experiment, we
therefore must average over multiple experiments in different gauges
to get the true sampling behavior. Because of these control errors,
it is not necessarily true that $\{P^{\textrm{low}}\}\rightarrow\{0,1\}$
even in the limit $n_{run}\rightarrow\infty$.}

\textcolor{black}{In general, a single spin can undergo many spin-sign
transitions. If we assume that a sufficient number of experiments
have been performed to be statistically confident in the experimental
result, $P_{i}^{\textrm{low}}{\color{red}{\color{black}<}}0.5$ implies
that $\sigma_{n_{run}}^{\textrm{experiment}}=\sigma_{i}^{\textrm{low}}$
. This does not generally demonstrate that the experimental system
is at a low temperature, only that it has undergone an even number
of spin-sign transitions from $T=\epsilon$. Conversely $P_{i}^{\textrm{low}}>0.5$
implies an odd number of transitions.}

\textcolor{black}{Let us now consider how to apply the methods to
real experimental data. If we are at a low enough temperature that
we can neglect the possibility that a spin has multiple spin-sign
transitions below that temperature, and assume that $n_{i}^{\textrm{trans}}\in\{0,1\}\;\forall i$,
then experimental measurements of $P_{i}^{\textrm{low}}$ allow for
a convenient way to check if the lowest temperature decoding transition
has occurred. If $P_{i}^{\textrm{low}}>0.5$ than we can conclude
that the transition has happened, otherwise we can conclude that it
has not. Now consider experimentally measuring $P_{i}^{\textrm{low}}$
for each bit in a series of randomly generated Ising Hamiltonians,
and plotting the result as a function of the theoretically calculated
spin-sign transition temperature, $T_{trans}$. If the experimental
system produced a perfect Boltzmann distribution at a temperature
$T^{\textrm{experiment}}$, then in the limit $n_{run}\rightarrow\infty$
this plot would look like a step function going from one to zero centered
at $T^{\textrm{experiment}}$. Reducing $n_{run}$to a finite value
will broaden out this sharp transition, but will not affect the temperature
at which the plot crosses $P^{\textrm{low}}=0.5$. We can calculate
the temperature dependence of the spin orientations by explicitly
constructing the Boltzmann distribution. We can then apply the central
limit theorem to calculate the theoretically expected $P_{low}$:}

\textcolor{black}{
\[
P_{i}^{\textrm{low}}(T_{trans},n_{run})\equiv
\]
}

\textcolor{black}{
\begin{equation}
\frac{1}{2}\textrm{erfc}\left(2\,\left(\frac{1}{2}-\frac{\sum_{j}\sigma_{i}^{_{j}}\exp(-\frac{E_{j}}{T_{trans}})}{\sum_{j}\exp(-\frac{E_{j}}{T_{trans}})}\right)\,\sqrt{n_{run}}\right),\label{eq:plow_theory}
\end{equation}
}

\textcolor{black}{where $\sigma_{i}^{_{j}}\in\{1,-1\}$ is the orientation
of bit $i$ within state $j$.}

\textcolor{black}{Let us now consider the transitions for all single
cell Hamiltonians with $J_{ij}\in\pm1$ at some energy scale $\alpha$.
At first sight the calculation of all of these spin-sign transitions
temperatures seems like a monumental task given that naively there
should be $2^{N+M}=2^{24}$ different Hamiltonians. However, only
Hamiltonians which cannot be mapped into each other by gauge or symmetry
transformations can have different spin-sign transitions. The gauges
allow us to reduce the number of Hamiltonians we examine to $2^{16}$
while the symmetries of the unit cell allow us to further reduce this
number to $192$. We are therefore able to experimentally test each
of these Hamiltonians on the D-Wave chip, as described in the next
section. It is important that, in the case of the single unit cell,
there are no spins which show more than one spin-sign transition at
finite temperature. The transition plot should therefore be a valid
way to analyze the decoding at any temperature without having to consider
errors due to multiple transitions.}

\textcolor{black}{The next natural question concerns the performance
on larger systems. For this purpose, we consider a 4x4 chimera graph
with 128 qubits. While the calculation of the spin-sign transition
temperatures cannot be performed by exhaustive search the way the
single unit cell was, it can be efficiently solved using a bucket
tree elimination (BTE) \cite{Kask(2001)}. A software sampler based
on BTE allows us to estimate the orientation of each spin as a function
of temperature. From this estimate we can extract $\{T_{trans}\}_{i}$
and $\sigma_{i}^{\textrm{low}}$ for each bit. For a detailed explanation
of our methods see Sec. 4 of the supplemental material. For spins
with a ground state orientation of zero, the orientation curve as
a function of $T$ remains very close to zero at low temperatures.
This is problematic for our analysis because statistical error can
cause us to detect a large number of spurious transitions. For this
reason we exclude these data from our analysis.}

\textcolor{black}{We quantify the performance of the chip for maximum
entropy tasks on a larger system in two ways. First we can compare
all Hamiltonians with a Boltzmann distribution at the same temperature,
and define}

\textcolor{black}{
\begin{equation}
P_{err}(T)=\sum_{H\in\textrm{Hamiltonians}}P_{err}^{H}(T)/N_{H}
\end{equation}
}

\textcolor{black}{where $N_{H}$ is the number of Hamiltonians examined
and}

\textcolor{black}{
\begin{equation}
P_{err}^{H}(T)=\sum_{i=1}^{N_{trans}}|\sigma_{i}^{\textrm{decoded}}(T)-\sigma_{i,\;\textrm{total}}^{\textrm{experiment}}|/(2\, N_{trans}).
\end{equation}
}

\textcolor{black}{Second we admit the possibilities that different
Hamiltonians freeze at different times and examine $\min_{T}(P_{err}^{H}(T))$
for each Hamiltonian individually. It is important to emphasize again
that we only include bits with at least one spin-sign transition in
this analysis to avoid the data being overwhelmed by spins which behave
trivially.}

\subsection*{\textcolor{black}{Microscopic results}}

\begin{figure}
\begin{centering}
\textcolor{black}{\includegraphics[scale=0.5]{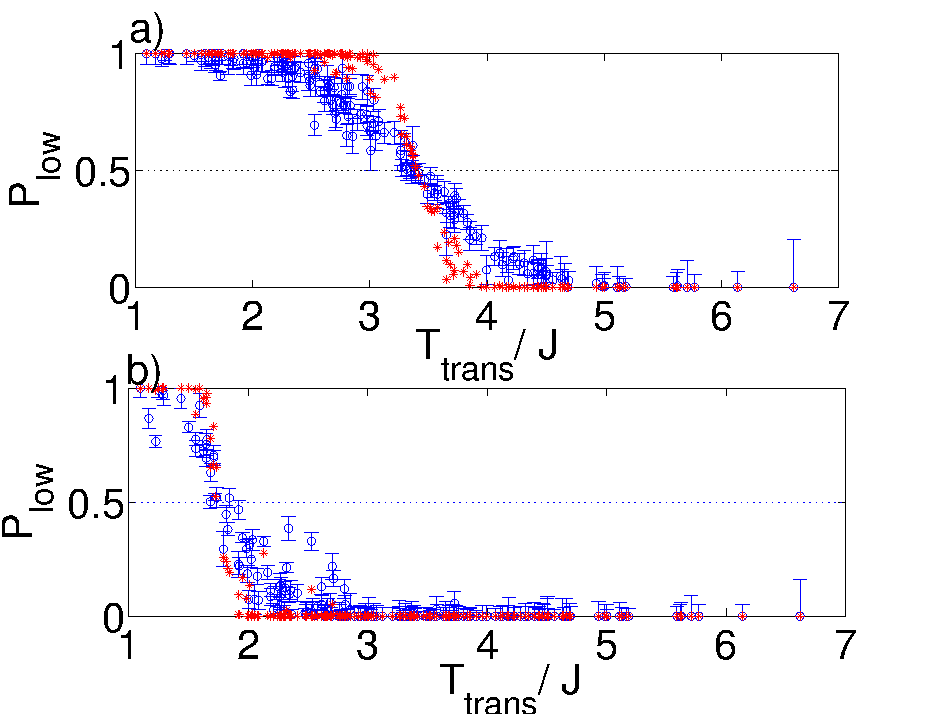} }
\par\end{centering}

\textcolor{black}{\caption{\label{fig:single_uc_indv}Single unit cell decoding transitions.
(color online) Circles with errorbars (blue) are experimentally-measured
values of the probability of each bit decoding to the predicted low-temperature
result plotted as a function of the calculated spin-sign transition
temperature. Error bars show one standard deviation. Asterisks (red)
are the calculated behavior for a Boltzmann distribution with a temperature
determined by fitting to Eq. \ref{eq:plow_theory}. Dotted line at
$0.5$ is a guide for the eye. a) $\alpha=0.05$; in this case the
fitted temperature is $T_{fit}=0.170\, J$. b) $\alpha=0.1$; in this
case the fitted temperature is $T_{fit}=0.175\, J$. }
}
\end{figure}

\textcolor{black}{In Fig. \ref{fig:single_uc_indv} we show the experimentally-measured
probability that each bit in a single Chimera unit cell decodes to
the calculated low temperature orientation, plotted as a function
of the calculated spin-sign transition temperature. We also show the
theoretical dependence obtained by assuming that the annealing device
precisely follows a Boltzmann distribution - i.e. equation \ref{eq:plow_theory}.
The data plotted in Fig. \ref{fig:single_uc_indv} demonstrate that
for this example the device should perform maximum entropy tasks quite
well. However the data differ from what would be expected from a device
which samples a pure Boltzmann distribution: in both plots the transition
is significantly broader than the pure Boltzmann result. This effect
can most likely be attributed to control errors, which cause random
deviations in the couplers and the fields. It is also worth noting
that especially in \ref{fig:single_uc_indv} b) there are some individual
transitions which decode correctly but deviate strongly from what
we would expect from a Boltzmann sampler. These could be due to dynamical
effects similar to the ones explored in \cite{Boixo(2013)}. For this
reason the Hamiltonians associated with these transitions may warrant
future study. In the interest of future work, we include these Hamiltonians
in Sec. 5 of the Supplemental Material.}

\textcolor{black}{}

\textcolor{black}{The fits performed for the single unit cell can
provide us an estimate of a phenomenological temperature for the device
which seems quite robust. We can use this effective temperature combined
with the known annealing schedule and physical temperature of the
device to find the parameters at the 'freeze time' when the dynamics
effectively stop. Tab. \ref{tab:single_uc_freeze} displays these
results. As one may intuitively expect, the system appears to freeze
earlier for larger values of $\alpha$. Although the difference in
apparent temperature is relatively small, it corresponds to freezing
with a much higher value of transverse field. We note however that
in all cases the transverse field at the freeze time is much weaker
than the couplers.}

\textcolor{black}{}
\begin{table*}
\begin{centering}
\textcolor{black}{}%
\begin{tabular}{|c|c|c|c|c|}
\hline 
\textcolor{black}{$\alpha$ } & \textcolor{black}{$\frac{T_{fit}}{\alpha\, B_{freeze}}$ } & \textcolor{black}{$B{}_{freeze}$ GHz (mK) } & \textcolor{black}{$\frac{t_{freeze}}{t_{f}}$ } & \textcolor{black}{$A_{freeze}$ GHz (mK)}\tabularnewline
\hline 
\hline 
\textcolor{black}{$0.05$ } & \textcolor{black}{$0.170$ } & \textcolor{black}{$2.08$ ($99.8$) } & \textcolor{black}{$0.748$ } & \textcolor{black}{$0.0453$ ($2.17$)}\tabularnewline
\hline 
\textcolor{black}{$0.1$ } & \textcolor{black}{$0.175$ } & \textcolor{black}{$2.02$ ($96.9$) } & \textcolor{black}{$0.733$ } & \textcolor{black}{$0.0530$ ($2.54$)}\tabularnewline
\hline 
\textcolor{black}{$0.15$ } & \textcolor{black}{$0.180$ } & \textcolor{black}{$1.86$ ($89.3$) } & \textcolor{black}{$0.696$ } & \textcolor{black}{$0.0844$ ($4.05$)}\tabularnewline
\hline 
\end{tabular}
\par\end{centering}

\textcolor{black}{\caption{\label{tab:single_uc_freeze}Table of effective temperatures and corresponding
parameters at the freeze time. By fitting experimental data to Eq.
\ref{eq:plow_theory} we can extract the ratio of the temperature
to the strength of the couplings $\frac{T_{fit}}{\alpha\, B_{freeze}}$.
Because the temperature is fixed during the annealing process and
the coupling changes monotonically with a known annealing schedule
we are able to extract the value of the Ising energy scale $B_{freeze}$
at which the system can no longer equilibrate. We then use the known
annealing schedule to extract the freeze time, $t_{freeze}$, and
the transverse field energy scale at the freeze time, $A_{freeze}$.
The device base temperature is $17$ mK.}
}
\end{table*}

\begin{figure}
\begin{centering}
\textcolor{black}{\includegraphics[scale=0.5]{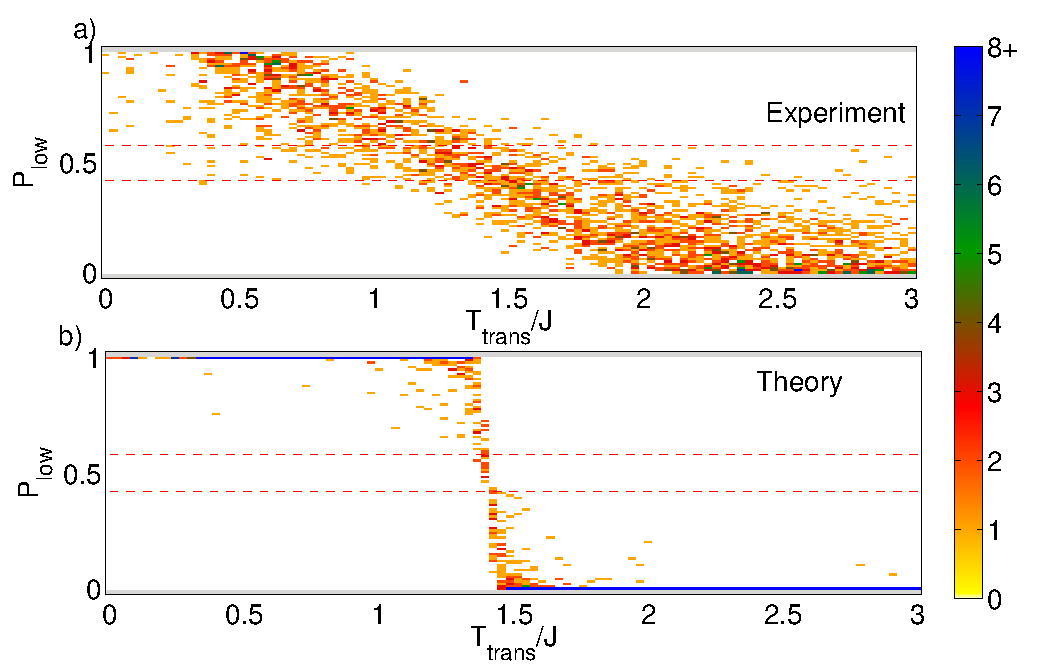} }
\par\end{centering}

\textcolor{black}{\caption{\label{fig:4x4_indv_p15} Densities shown in (a) are counts for experimentally-measured
probabilities of decoding to low temperature result on the y-axis
and theoretically determined transition temperatures on the x-axis.
(b) is the same, but where the values on the y-axis are determined
theoretically using Eq. \ref{eq:plow_theory} and $T/(\alpha\, J)=1.405$.
Both plots are for a 4x4 chimera with $\alpha=0.15$. Data are based
on $145$ randomly chosen Hamiltonians with $200$ flipped bonds or
fields. For each Hamiltonian $100$ sets of $1000$ annealing cycles
are performed. $P_{low}$ is the fraction of these sets of annealing
cycles for which the spin agrees with the low temperature result.
Dashed (red) lines are the distance from $P_{low}=0.5$ where data
become statistically significant at the $95\%$ level. Note that this
density plot is only for spins with a single $T_{trans}$; spins with
multiple transitions have been observed but are excluded from these
figures. }
}
\end{figure}

\textcolor{black}{Unlike the case of the single unit cell, we cannot
possibly exhaustively sum over all inequivalent Hamiltonians for a
4x4 unit cell block. We can however randomly sample Hamiltonians and
look at the transitions. Fig. \ref{fig:4x4_indv_p15} a) displays
such a plot. As before we can compare with what would be expected
for a machine which samples an ideal Boltzmann distribution (with
the same number of samples per run) shown in, Fig. \ref{fig:4x4_indv_p15}
b). From comparing \ref{fig:4x4_indv_p15} a) and b) we observe that
the step function shape is broadened much more than expected from
finite sample size alone. The broadening is likely due to control
errors (i.e. errors in the specification of $h_{i}$and $J_{ij}$
on the chip), which are different in each run and therefore would
be expected to create broadening even if the number of samples per
run is large. This is confirmed by simulations in which random control
errors are added to the fields and couplers as shown in figure 8(a). }

\textcolor{black}{}

\begin{figure}
\begin{centering}
\textcolor{black}{\includegraphics[scale=0.5]{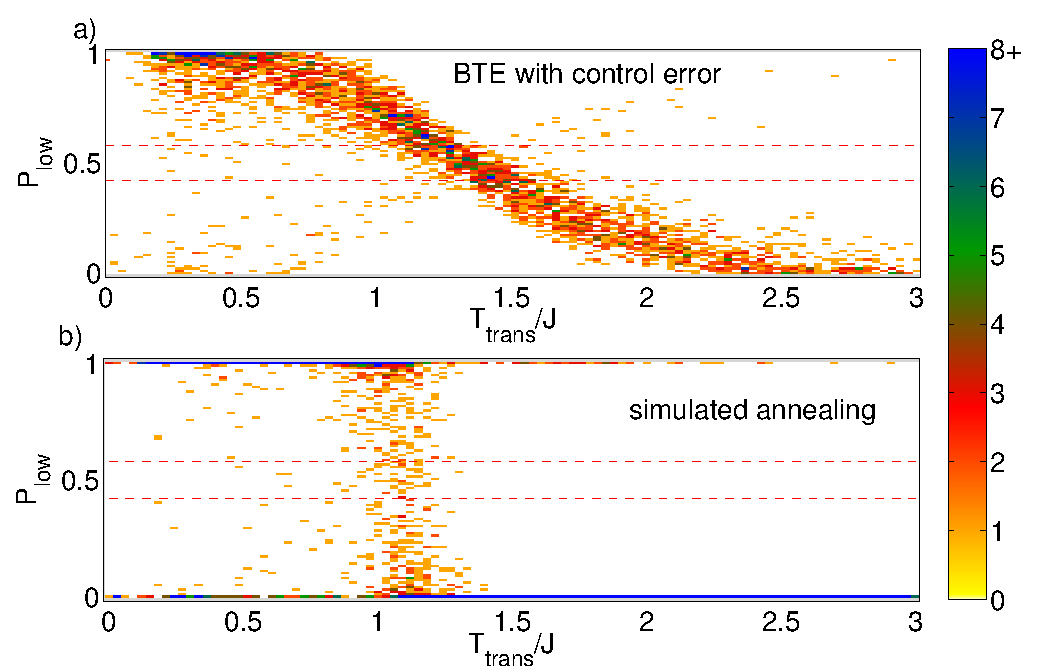}}
\par\end{centering}

\textcolor{black}{\caption{\textcolor{black}{\label{fig:err_equil_fail}Top: bucket tree elimination
sampled over independent control error of 5\% J in the fields and
3\% J in the couplers. These data represent perfect equilibration
subject to control error. Bottom: Simulated Annealing (SA) with 10,000
total updates and no control error. These data were taken with a linear
sweep starting from $T/(\alpha J)=10.$ The error in these data come
from failure to equilibrate. These data were calculated based on 1000
samples per run as were used experimentally. Dashed (red) lines are
the distance from $P_{low}=0.5$ where data become statistically significant
at the $95\%$ level.}}
}
\end{figure}

\textcolor{black}{In figure 8(b) we show the results of simulated
annealing in the absence of any control errors. By limiting the simulation
to a total of 10,000 updates we see that failure to equilibrate produces
a signature in which spins consistently decode with an orientation
different from the one expected from Boltzmann statistics at the temperature
which the system is operated. We do not observe this signature experimentally,
which suggests that the device is in fact reaching equilibrium, which
is consistent with the conclusions of \cite{Katzgraber2014}. Our
experimental data do however display a clear signature of control
error, which manifests itself in a broadening of the transition between
low and high temperature decoding. As Fig. \ref{fig:err_equil_fail}a)
demonstrates, this signature can be reproduced using BTE to calculate
perfect equilibration in the presence of realistic control error.
For further discussion comparing BTE and SA results, see Sec. 6 of
the supplemental material accompanying this paper. The astute reader
will note that Fig. \ref{fig:err_equil_fail}b) demonstrates a transition
which is apparently shifted to a lower temperture. We suspect that
this is due to the character of the local minima in which this simulation
becomes trapped corresponding to an effectively lower temperature.
It would be interesting to investigate this feature further, but such
an investigation is beyond the scope of this work.}

\textcolor{black}{}

\textcolor{black}{We now want to compare the data used to create Fig.
\ref{fig:4x4_indv_p15} directly with those expected for an ideal
Boltzmann distribution. For a given temperature and a given Hamiltonian
we can calculate the expected decoding with the Boltzmann distribution
and divide each bit with a transition into two categories, the ones
which should decode to $\sigma^{\textrm{low}}$and those which should
not, or equivalently those which have undergone even and odd numbers
of transitions (including 0) from $T=\epsilon$. Only some of the
experimental data are statistically significant. We consider only
which are far enough away from $P_{low}=0.5$.}

\textcolor{black}{}
\begin{figure}
\begin{centering}
\textcolor{black}{\includegraphics[scale=0.5]{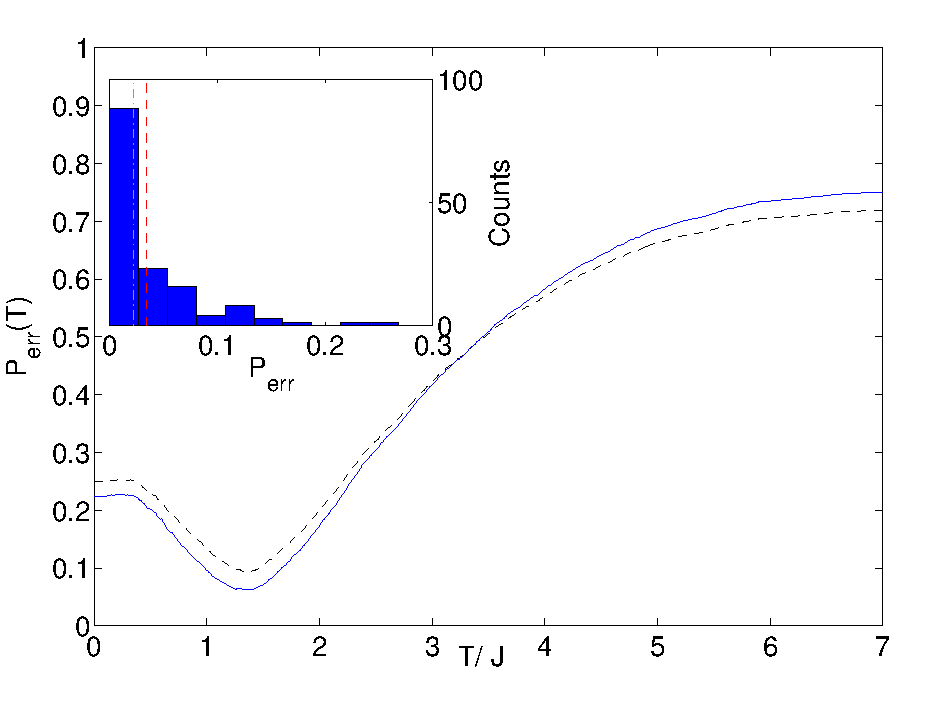} }
\par\end{centering}

\textcolor{black}{\caption{\label{fig:4x4_indvT}\textcolor{black}{The main plot depicts the
probability that the experiment finds an erroneous result for decoding
a given spin assuming the underlying values of T/J given on the x-axis.
Note that this is the s}ame data set as Fig. \ref{fig:4x4_indv_p15}
(a) but including spins with multiple transitions . Rate of decoding
errors versus temperature, dashed (black) line is all data, while
solid (blue) line represents only the data which are statistically
significant at the $95\%$ level. \textcolor{black}{Inset:} Histogram
of error rates at the best temperature for each Hamiltonian, $\min_{T}(P_{err}^{H}(T))$.
Dot-dashed lines are median, dashed lines are mean. A summary of relevant
quantities extracted from these data can be found in Tab. \ref{tab:4x4_stats}. }
}
\end{figure}

\textcolor{black}{We plot $P_{err}(T)$ in Fig. \ref{fig:4x4_indvT}.
It shows a temperature for optimum performance which roughly agrees
with our analysis of the single unit cell. A simple way to analyze
the performance is to find the single temperature where the total
performance summed over all of the Hamiltonians is the best. An alternate
approach is to allow for the fact that different Hamiltonians may
become 'frozen' at different points near the end of the annealing
process and therefore different temperatures. Under this approach,
$P_{err}$ should be minimized for each Hamiltonian individually.
A histogram of these data appears in the inset. From this histogram
we notice that the device appears to perform quite well typically,
but occasionally performs rather poorly.}

\textcolor{black}{We list relevant performance metrics for different
values of $\alpha$ in Tab. \ref{tab:4x4_stats}. The table shows
better performance by all metrics for larger values of $\alpha$,
again consistent with control errors being the limiting factor since
at smaller $\alpha$, the control errors will be more significant
compared to the energy scale of the Hamiltonian. On the other hand,
if the performance were limited because the dynamics were unable to
reach equilibrium than we would expect to see the opposite trend,
as smaller energy barriers would mean that the system is more able
to reach equilibrium.}

\textcolor{black}{}
\begin{table}
\begin{centering}
\textcolor{black}{}%
\begin{tabular}{|c|c|c|c|c|}
\hline 
 & \multicolumn{2}{c|||}{\textcolor{black}{same T}} & \multicolumn{2}{c|||}{\textcolor{black}{diff. T $95\%$sig.}}\tabularnewline
\hline 
\hline 
\textcolor{black}{$\alpha$ } & \textcolor{black}{overall } & \textcolor{black}{$95\%$ sig. } & \textcolor{black}{mean } & \textcolor{black}{median }\tabularnewline
\hline 
\textcolor{black}{$0.05$ } & \textcolor{black}{$14.69\%$ } & \textcolor{black}{$9.30\%$ } & \textcolor{black}{$5.78\%$ } & \textcolor{black}{$3.03\%$ }\tabularnewline
\hline 
\textcolor{black}{$0.1$ } & \textcolor{black}{$11.02\%$ } & \textcolor{black}{$8.15\%$ } & \textcolor{black}{$5.07\%$ } & \textcolor{black}{$2.82\%$ }\tabularnewline
\hline 
\textcolor{black}{$0.15$ } & \textcolor{black}{$9.30\%$ } & \textcolor{black}{$6.23\%$ } & \textcolor{black}{$3.48\%$ } & \textcolor{black}{$2.22\%$ }\tabularnewline
\hline 
\end{tabular}
\par\end{centering}

\textcolor{black}{\caption{\label{tab:4x4_stats} Different metrics for the error rates at different
values of $\alpha$. Same T indicates minimisation with the constraint
that all Hamiltonians be modeled with the same effective temperature,
while diff. T indicates that each one is allowed a different temperature.
The data for 'same T' with $\alpha=0.15$ are the minima of the two
curves in Fig. \ref{fig:4x4_indvT} (main figure) while the 'diff.
T' data are the mean and median illustrated as the red and green line
respectively in the inset of the same figure. }
}
\end{table}

\section*{\textcolor{black}{Conclusion}}

\textcolor{black}{We have shown that maximum entropy approaches to
decoding, implemented using the D-Wave chip programmable Josephson
junction annealer, can result in somewhat improved accuracy. This
confirms that useful information can be extracted from the excited
states (which, in any real machine at finite temperature, will have
non-zero occupation) at the end of the annealing process. Our results
are applicable to a wide range of problems in machine learning and
optimization provided knowledge of prior information. Applications
for which the maximum likelihood method is known to perform poorly
or those where marginal benefits are valuable would be most suited
to the current D-Wave architecture which features two-local couplers.
An extension of the architecture to allow $m$-local couplers (with
$m\geq3$) would allow the Shannon limit to be approached in decoding
applications\cite{Kashimba(2004)}.}

\textcolor{black}{We further show using analysis on the individual
spin level that the experimental device produces a distribution which
is Boltzmann like and therefore suitable for maximum entropy tasks.
Additionally we demonstrate that the most probable limiting factors
for such tasks are control errors, rather than equilibration dynamics.
While our work has focused on the exploitation of the D-wave processor
as a classical thermal annealing device, more experiments (e.g. by
varying the annealing schedule) to determine the role (if any) of
quantum fluctuations on maximum entropy inference using the D-Wave
processor would be illuminating. We also note that generalizations
of the methods we have outlined here may be useful to study disordered
magnets such as spin glasses.}

\subsection*{\textcolor{black}{Author Contributions}}

\textcolor{black}{NC performed the experiments for the published data,
while WV performed early experiments and data analysis which guided
the direction of the project. NC, SS, and PW performed the data analysis
for the published data. NC, PW, and GA wrote the paper. All authors
were involved in discussing the results and editing the manuscript.}

\subsection*{\textcolor{black}{Competing Financial Interests}}

\textcolor{black}{The authors declare no competing financial interests.}

\subsection*{\textcolor{black}{Acknowledgements}}

\textcolor{black}{This work was supported by Lockheed Martin and by
EPSRC (grant refs: EP/K004506/1 and EP/H005544/1). We thank the USC
Lockheed Martin Quantum Computing Center at the University of Southern
California's Information Sciences Institute for access to their D-Wave
Two machine. We acknowledge fruitful discussions with Mike Davies,
David Saad, Andrew Green, and Greg Ver Steeg. The BTE solver was provided
by D-Wave Systems Inc. as part of the chip operation API. }

\part*{Supplemental Material for: Maximum-Entropy Inference with a Programmable
Annealer}

\author{Nicholas Chancellor{*}'', Szilard Szoke\textasciicircum{}, Walter
Vinci'$^{\dagger}$,\\
 \textasciicircum{} Gabriel Aeppli$^{\ddagger}$ and Paul A. Warburton''\textasciicircum{}}

\lyxaddress{``London Centre For Nanotechnology 19 Gordon St, London UK WC1H
0AH\\
 \textasciicircum{}Department of Electronic and Electrical Engineering,
UCL, Torrington Place London UK WC1E 7JE\\
 ' University of Southern California Department of Electrical Engineering
825 Bloom Walk Los Angeles CA. USA 90089 \\
 $^{\dagger}$University of Southern California Center for Quantum
Information Science Technology 825 Bloom Walk Los Angeles CA. USA
90089\\
 $^{\ddagger}$Department of Physics, ETH Zürich, CH-8093 Zürich,
Switzerland\\
 Department of Physics, École Polytechnique Fédérale de Lausanne (EPFL),
CH-1015 Lausanne, Switzerland\\
 Synchrotron and Nanotechnology Department, Paul Scherrer Institute,
CH-5232, Villigen, Switzerland}

\section{Calculation of Continuous decoded bit-error-rate Curves}

Naively, it would appear that to plot message bit-error-rate versus
crossover probability curves we would need to take a statistically
significant sample for each value of $T_{Nish}$. However in this
section we demonstrate that this is in fact not the case, by a careful
choice of how we collect and analyze the experimental data, we can
efficiently calculate the final bit-error-rate for \emph{any} value
of $T_{Nish}.$ This method allows us to plot continuous curves, as
displayed in Fig. 4 of the main text.

Let us first consider the general form of the Hamiltonian for our
Ising system $H=\alpha(-\sum_{i}h_{i}\sigma_{i}^{z}-\sum_{i,j\in\chi}J_{ij}\sigma_{i}^{z}\sigma_{j}^{z})$
, subject to the constraint that $h_{i}\in\{-1,1\}$and $J_{ij}\in\{-1,1\}$
. Due to this constrain there are a finite number of Hamiltonians
which can be generated by the noisy channel, let us call this set
$\{H\}$. For each $H_{n}\in\{H\}$we can count the number of corrupted
bits in the recieved codeword $N_{corr}^{(n)}\equiv\sum_{i}\delta_{h_{i}^{(n)},-1}+\sum_{i,j\in\chi}\delta_{J_{ij}^{(n)},-1}$.
Because all bits are subject to the same crossover probability, $p$,
the total probability of having a given number of bits flipped can
be written as 

\begin{equation}
q(H_{n},p)\equiv q(N_{corr},p)=\frac{1}{2^{N+M}}(p)^{N_{corr}}(1-p)^{N+M-N_{corr}}\tbinom{N+M}{N_{corr}},
\end{equation}

where $N+M$ is the total number of elements (couplers and fields)
in the Hamiltonian. We can assign to each Hamiltonian a bit error
rate $r_{n}$, for the purposes of this discussion $\{r\}$ could
have either been calculated (for example by exhaustive summing or
BTE) or obtained experimentally. We now write the total bit-error-rate
as a function of crossover probability 

\begin{equation}
r_{tot}(p)=\sum_{n=1}^{2^{N+M}}q(N_{corr}^{(n)},p)\, r_{n}.\label{eq:r_ungroup}
\end{equation}

We now observe that$N_{corr}\in\{0,1,2\ldots N+M\}$ so for a given
value of $p$ there are only $N+M+1$ possible values of $p'(N_{corr},p)$.
Based on this observation we can group these terms together to write
down the total decoded bit-error-rate
\begin{equation}
r_{tot}(p)=\sum_{s=0}^{N}q(s,p)\dbinom{N+M}{s}^{-1}\sum_{n'=1}^{\tbinom{N+M}{s}}r_{s,n'}=\sum_{s=0}^{N}q(s,p)\,\bar{r}_{s},\label{eq:r_group}
\end{equation}

where $r_{s,n'}$is the list of bit-error-rates for all Hamiltonians
with $N_{corr}=s$ and $N'_{s}$ is the number of Hamiltonians sampled
in sector $s$. Based on the formula given in Eq. \ref{eq:r_group}we
can make several observations. Firstly we not that $\bar{r}_{s}$
does not depend on $p$, and $q$ does not depend on any of the decoding
rates, therefore knowing the $N+M+1$ values of$\{\bar{r}\}$ allows
us to easily calculate $r_{tot}(p)$ for any value of $p$. Secondly,
we note that $\bar{r}_{s}$ need not be an exhaustive sum over all
possible Hamiltonians with $s$ corrupted elements; a good approximation
of $r_{tot}(p)$ could be obtained using $\bar{r}_{s}$extracted from
a representative sample.

\section{Confirmation that $T_{Nish}=T$ is the optimal decoding temperature
for our theoretical data}

It is not immediately obvious that Fig. 3 obeys the famous result
by Nishimori that optimal decoding happens at $T_{Nish}=T.$ For this
reason we have plotted several slices of that figure in the $T$ direction
along with the value along the line where $T_{Nish}=T$ in Fig. S1.
We immediately note that in that plot the curve representing $T_{Nish}=T$
crosses the other curves at there gloabal minima, thereby confirming
that our theoretical data do in fact agree with this result. It is
worth noting however that the converse of this famous result is not
necessarily true, the minimum in the $T_{Nish}$ direction does not
necessarily occur at $T_{Nish}=T$.This arises from the fact that
the decoding curves are discontinuous in the $T$ direction. 

\begin{figure}
\begin{centering}
\includegraphics[scale=0.5]{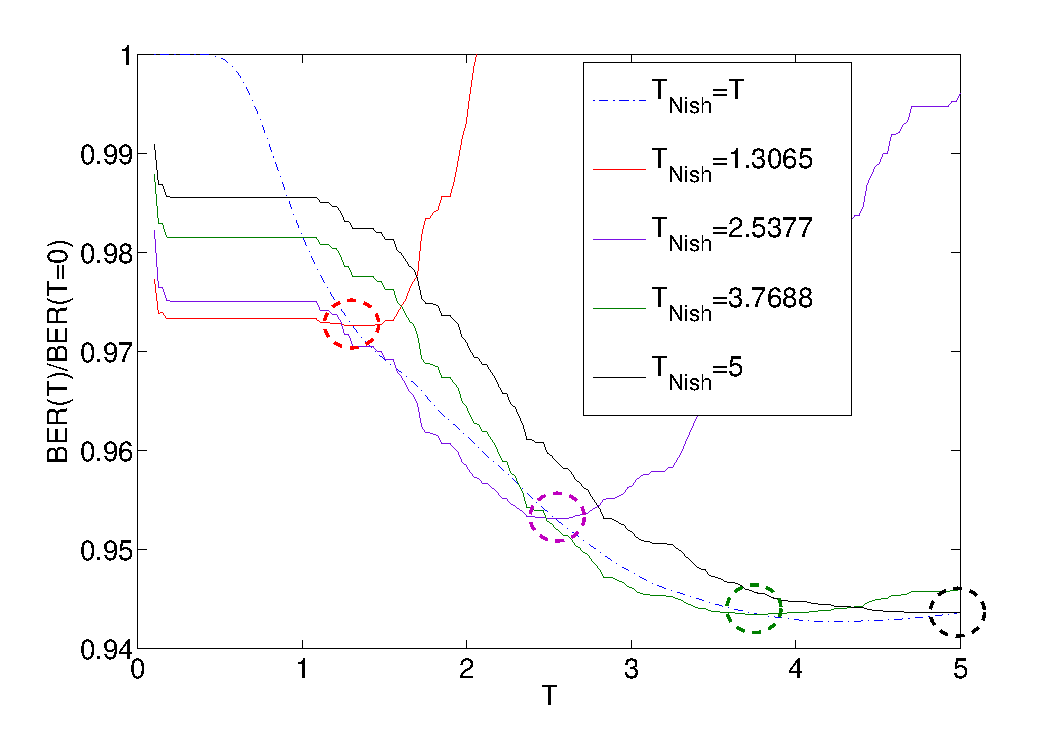}
\par\end{centering}

Figure S1: Theoretically predicted ratio of finite temperature bit-error-rate
to zero temperature bit-error-rate versus T for a variety of Nishimori
temperatures. Crossings at minimum values are circled for clarity.
Here BER is shorthand for bit-error-rate.
\end{figure}

\section{Effect of system size on decoding}

As we discussed in the main text, the plateau which can be seen in
Figs. 3 and 5 of the main text shinks and the discontinuities in the
ratio plotted in Fig. 3 decrease as the system size grows. The former
can be seen by examining the number of expected spin-sign transitions
for the 4x4 chimera versus the temperature at which the transition
occurs. As Fig. S2 demonstrates, there are clearly spin-sign transitions
within the 'plateau' region which was present for the single unit
cell. It is worth noting that transitions in this region are also
visible in Fig. S4. 

\begin{figure}
\begin{centering}
\includegraphics[scale=0.5]{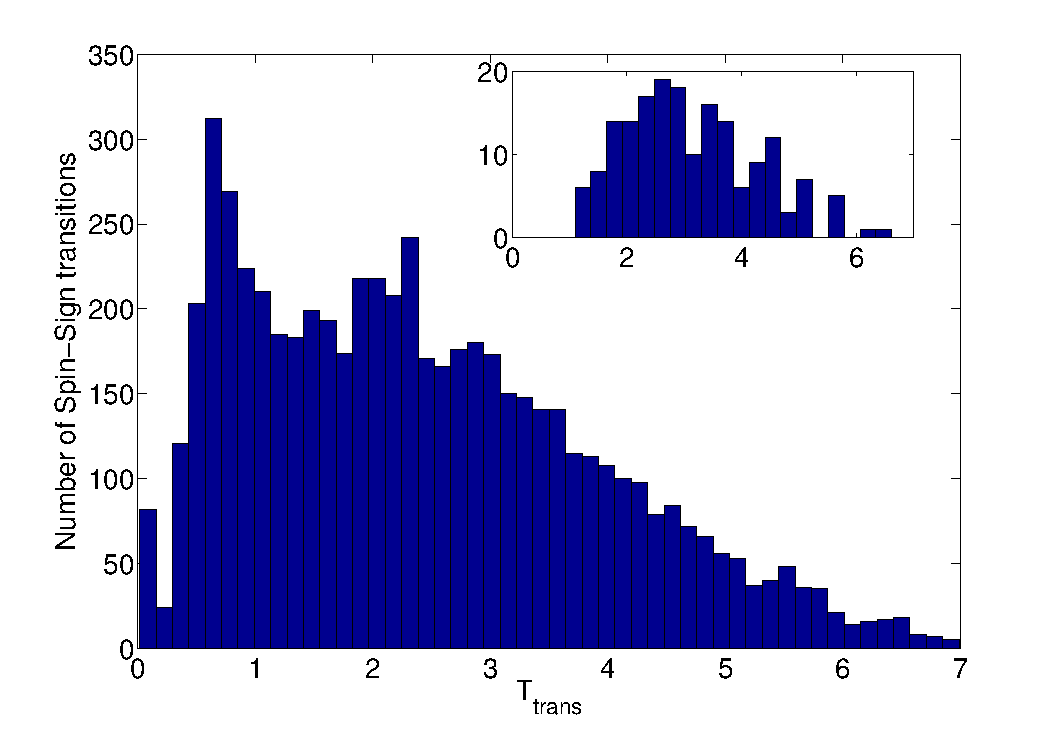}
\par\end{centering}

Figure S2 : Histogram of the number of spin-sign transitions versus
transition temperature for 145 Hamiltonians with 200 corrupted bits
each. Spins with a mean orientation of $0$ at $T=0$ have been excluded
to avoid plotting spurious transitions, as discussed in Sec. \ref{sec:Estimation-of-spin-sign}.
Inset: Same plot for single unit cell.
\end{figure}

To demonstrate the decrease in the size of the discontinuities with
system size, we produce the equivalent of Fig. 3 in the main text
but for a truncated version of the Chimera unit cell in which a bit
has been removed from each half of the bipartite graph. As Fig. S3
demonstrates, the discontinuities become much more severe. 
\begin{figure}
\begin{centering}
\includegraphics[scale=0.5]{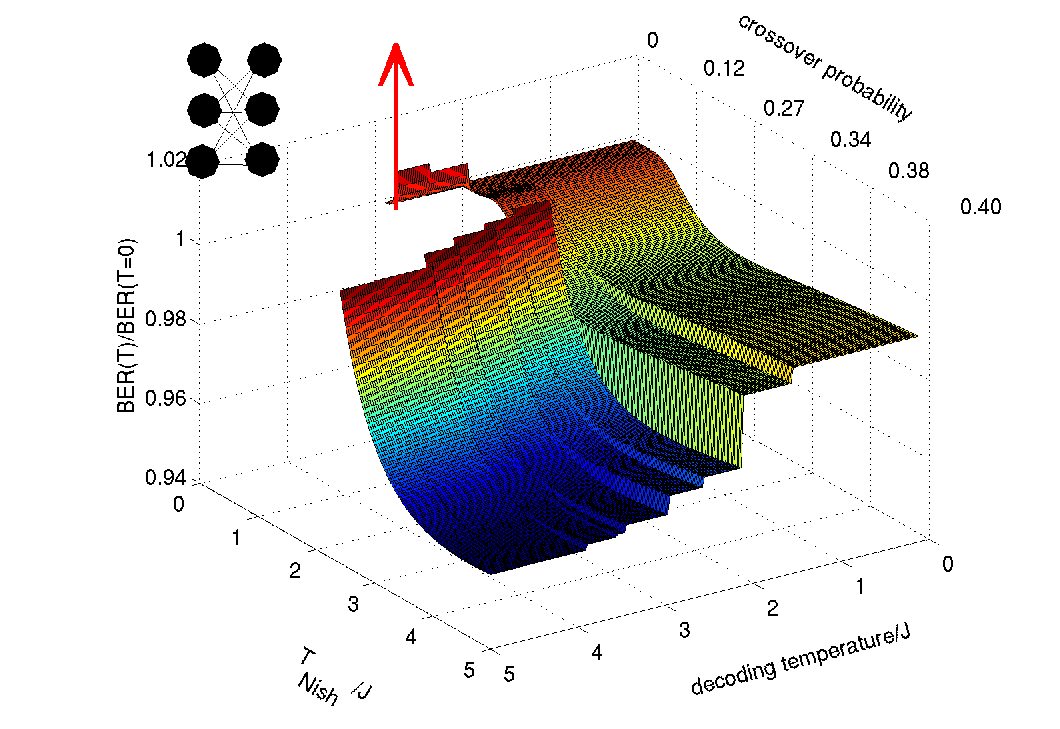}
\par\end{centering}

Figure S3: Analytically calculated bit-error-rate for a truncated
single unit-cell of the Chimera graph, plotted as a function of the
decoding temperature and the Nishimori temperature. The bit-error-rates
are normalized with respect to those obtained using maximum likelihood
decoding. The graph used for this analysis appears in the upper left
corner of the plot. Here BER is shorthand for bit-error-rate.
\end{figure}

\section{Estimation of spin-sign transitions for 4x4 Chimera\label{sec:Estimation-of-spin-sign}}

To estimate the temperatures at which spin-sign transitions occur,
we use Bucket Tree Elimination (BTE) code which has been provided
by D-Wave Systems Inc. This code acts as a fair Boltzmann sampler
and can be made to provide an unbiased sample of $N_{samp}$ spin
states at a given temperature $T$. We use $N_{samp}=10^{5}$. We
then sample 200 temperatures between $T=0$ and $T=7$ to determine
spin orientations. Fig. S4 gives an example of such data. 

\begin{figure}
\begin{centering}
\includegraphics[scale=0.5]{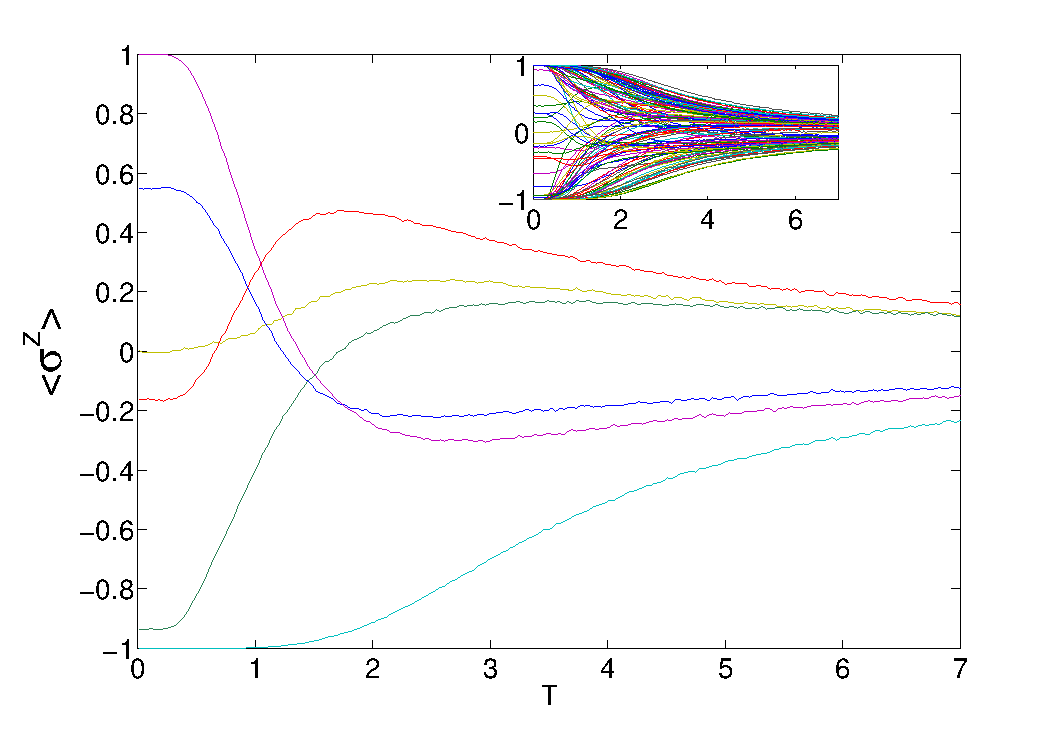}
\par\end{centering}

Figure S4: Example of theoretical orientation data for 4x4 chimera
unit cell Hamiltonian. Main Figure: selected spins. Inset: All 128.
\end{figure}

For each curve in Fig. S4 we can calculate the spin-sign transitions
by first taking a running average over 5 neighboring points to suppress
multiple spurious transitions and then using linear interpolation
to find the zero crossings. An example of this procedure is demonstrated
in Fig. S5. This method however becomes problematic when the orientation
remains very close to zero for a wide range of temperatures when the
ground state orientation is zero, as Fig. S6 demonstrates, statistical
sampling error leads to this method finding spurious transitions.
Because of these spurious transitions these spins are excluded from
our analysis.

\begin{figure}
\begin{centering}
\includegraphics[scale=0.5]{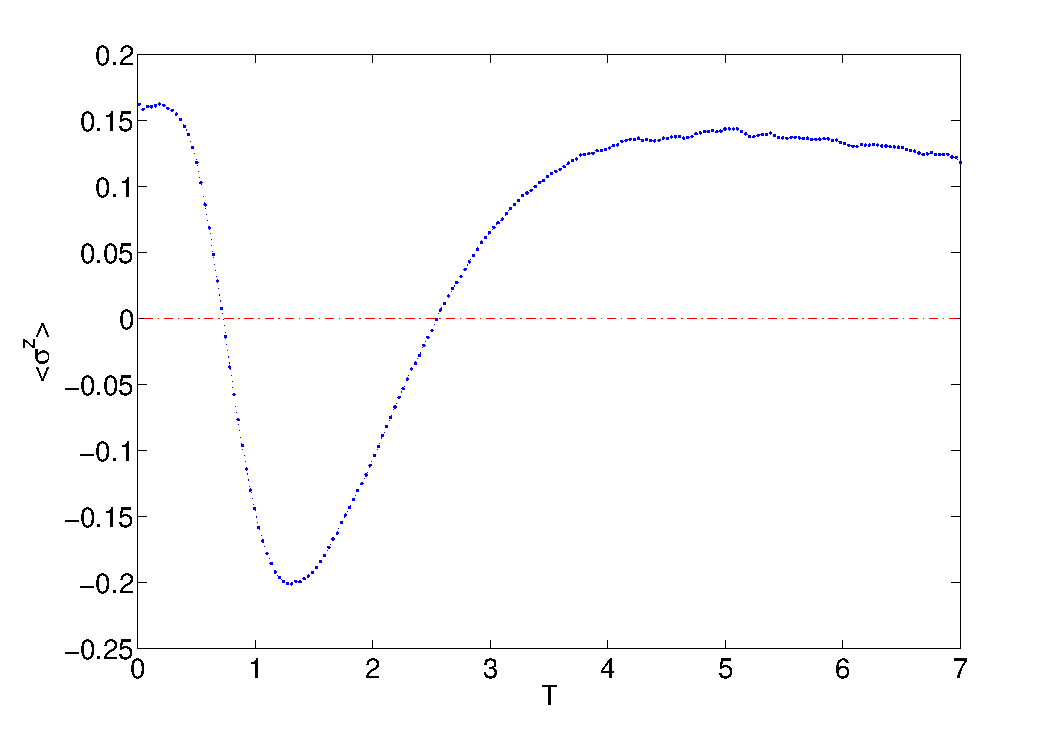}
\par\end{centering}

Figure S5: Example of finding spin-sign transitions using linear interpolation.
\end{figure}

\begin{figure}
\begin{centering}
\includegraphics[scale=0.5]{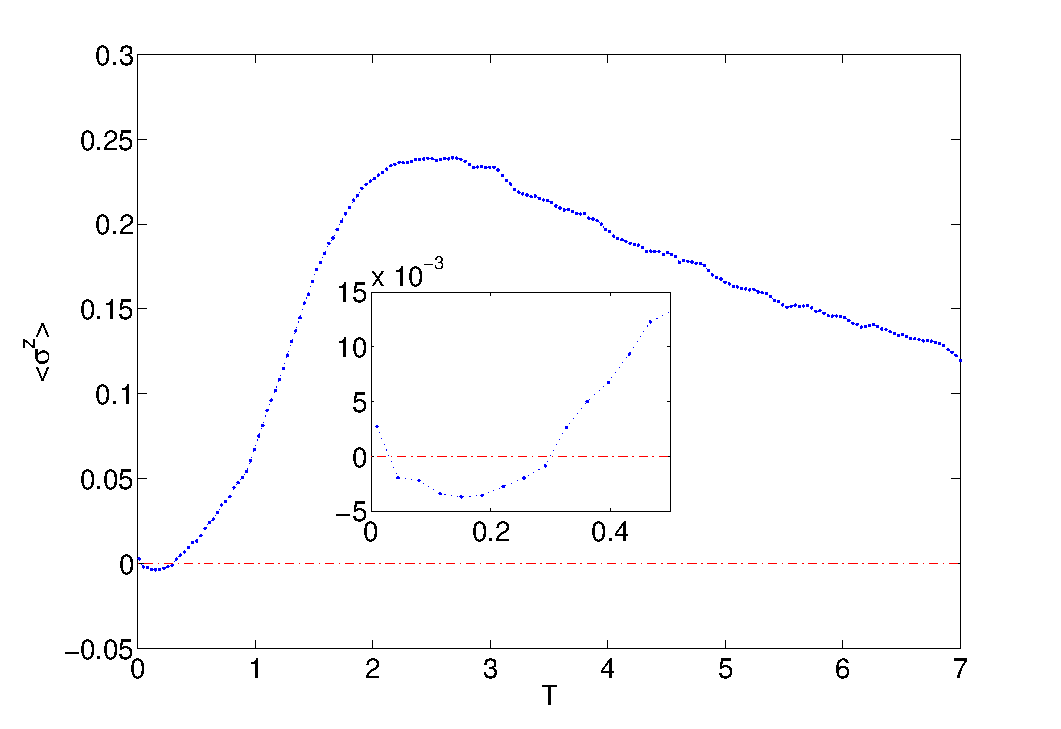}
\par\end{centering}

Figure S6: Example of spurious spin-sign transitions found when orientation
is zero at $T=0$, inset is zoom.
\end{figure}

\section{Hamiltonians which Decode Signifcantly Differently than Boltzmann}

We have not exhaustively searched for Hamiltonians which give significantly
different orientations from the Boltzmann distribution at many different
values of temperature, however a few have resulted as byproducts of
this project. In the interest of future work, we have included Fig.
S6 which displays 4 such Hamiltonians which we have found. Investigation
into the cause of these anomalies would be an interesting avenue of
future study.

\begin{figure}
\begin{centering}
\includegraphics[scale=0.5]{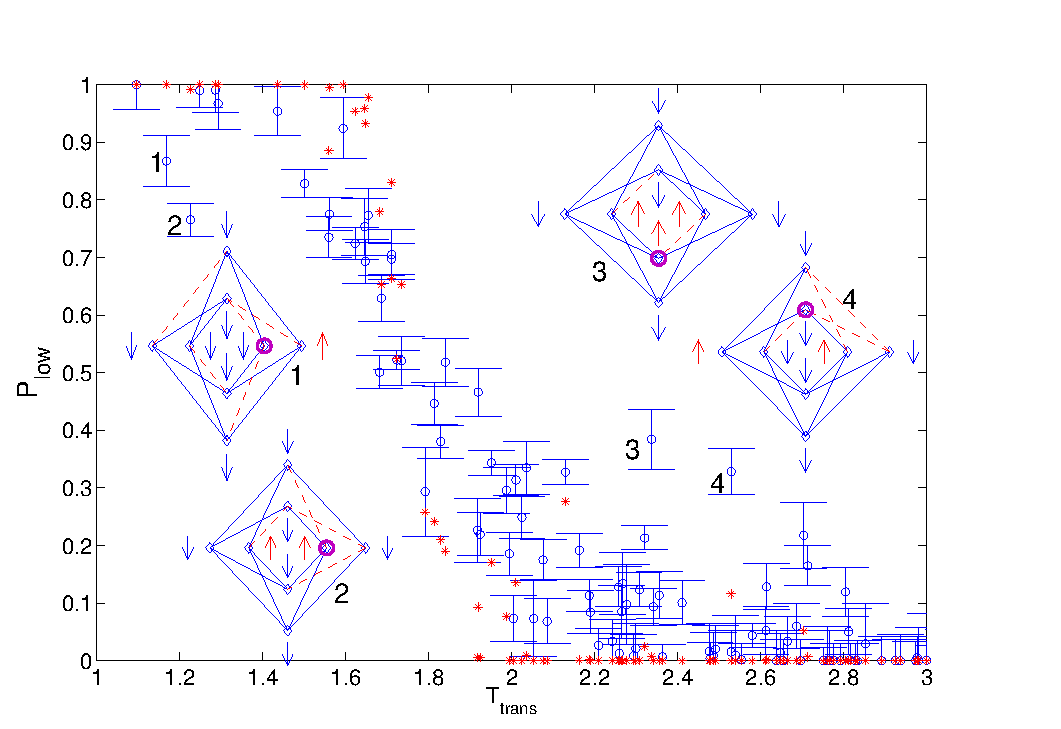}
\par\end{centering}

Figure S7: Spins with orientations which differ significantly from
Boltzmann distribution predictions. Each Hamiltonian is numbered and
the spin which demonstrates the anomalous decoding circled in purple.
Dashed lines are anti-ferromagnetic bonds, sold lines are ferro.
\end{figure}

\section{Comparison between Simulated Annealing and Bucket Tree Elimination
results}

\textcolor{black}{While BTE is guaranteed to provide an equilibrium
result, it is still interesting to ask whether we can achieve equilibrium
though simulated annealing. Fig. S8 demonstrates that for a randomly
selected 128 bit Hamiltonian from our data, a linear sweep from $T/(\alpha J)=10$
to $T/(\alpha J)=1.405$ can maintain equilibrium (which can be found
using BTE) throughout the entire sweep if 1,000,000 updates are used,
but fails to maintain equilibrium if only 10,000 are used, as was
the case in Fig. 8 of the main paper. }

\textcolor{red}{}
\begin{figure}
\begin{centering}
\textcolor{red}{\includegraphics[scale=0.5]{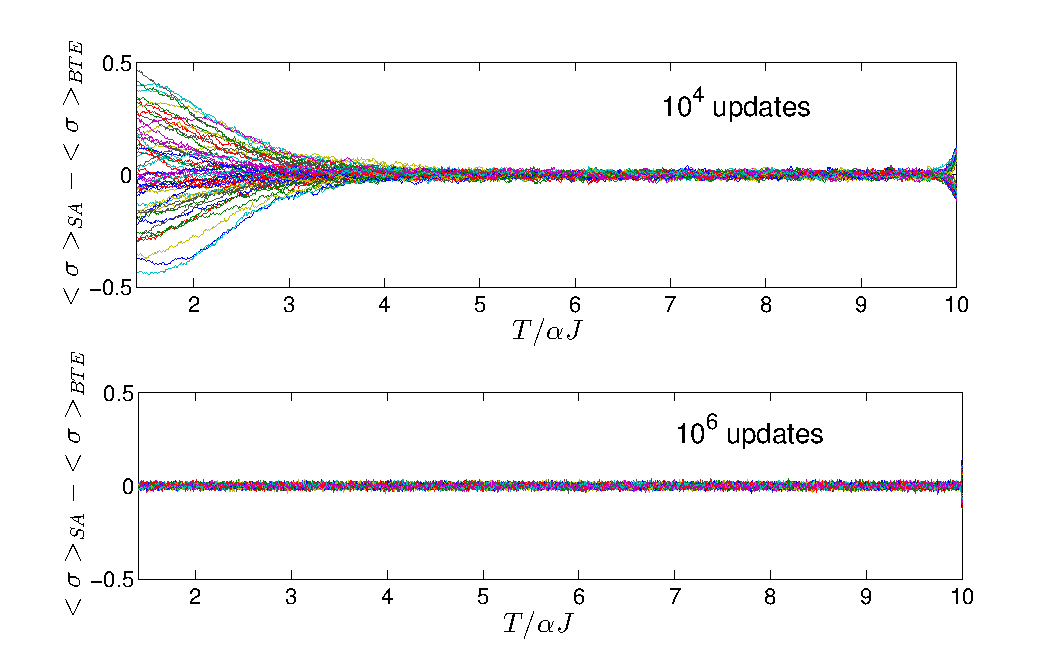}}
\par\end{centering}

\textcolor{black}{Figure S8: Difference between SA and BTE orientation
results for an SA run using a linear sweep from $T/(\alpha J)=10$
to $T/(\alpha J)=1.405$. These differences are plotted from all spins
with a spin-sign transition within one of the Hamiltonians used to
produce Fig. 7 of the main text, chosen at random. With only 10,000
updates the SA data begin to significantly deviate from thermal equilibrium
at $T/(\alpha J)\lessapprox3.5$, while for 1,000,000 updates the
data remain in equilibrium down to the temperature of interest, $T/(\alpha J)=1.405$.}
\end{figure}

\textcolor{black}{We further examine whether we can qualitatively
reproduce the behavior seen in Fig. 8a) of the main paper using SA
with many updates as well as control error. As Fig. S9 demonstrates,
we can reproduce this qualitative behavior. Slight differences between
Fig. S9 and 8a) of the main paper can be attributed either to not
all of the Hamiltonians fully equilibrating under SA, or to the fact
that restrictions on our available computing power have required us
to limit the number of samples used to produce Fig. S9.}

\begin{figure}
\begin{centering}
\includegraphics[scale=0.5]{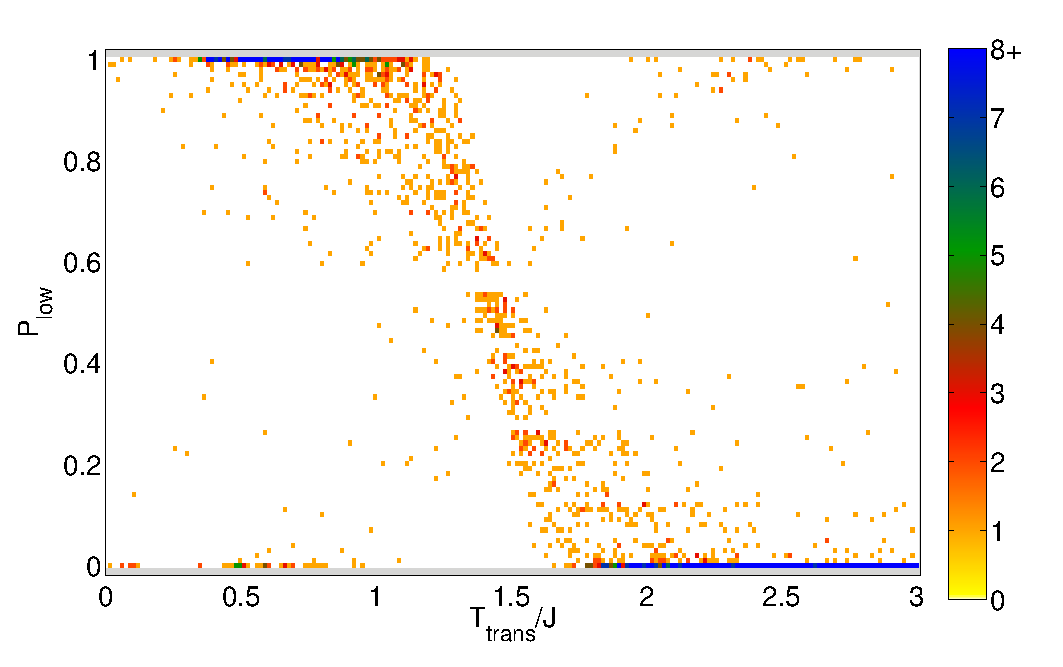}
\par\end{centering}

\textcolor{black}{Figure S9: Simulated Annealing data subject to control
error. These data were taken with a linear sweep starting from $T/(\alpha J)=5$
with $5\:10^{5}$ samples, 5\% field control errors and 3\% coupler
control error. Each randomly generated instance of the error was run
for 100 annealing runs and 100 instances were considered for each
Hamiltonian. }
\end{figure}

\end{document}